\documentclass[prx,twocolumn,amsmath,amssymb,floatfix,superscriptaddress]{revtex4-2}

\usepackage{graphicx}
\usepackage{dcolumn}
\usepackage{bm}
\usepackage{booktabs}

\usepackage{lipsum}

\usepackage[FIGTOPCAP,tight,raggedright,nooneline]{subfigure}
\usepackage{float}
\graphicspath{{images/}{plots/}}

\renewcommand{\vec}[1]{\bm{#1}}

\newcommand{\avg}[1]{\ensuremath{\left< #1 \right>}}

\newcommand{\abs}[1]{\ensuremath{\left\vert#1\right\vert}}

\DeclareMathOperator*{\argmax}{arg\,max}

\begin{document}

    \title{The bridges to consensus: Network effects in a bounded confidence opinion dynamics model}

    \author{Hendrik Schawe}
    \email{hendrik.schawe@cyu.fr}
    \affiliation{Laboratoire de Physique Th\'{e}orique et Mod\'{e}lisation, UMR-8089 CNRS, CY Cergy Paris Universit\'{e}, France}
    \author{Sylvain Fontaine}
    \affiliation{Laboratoire de Physique Th\'{e}orique et Mod\'{e}lisation, UMR-8089 CNRS, CY Cergy Paris Universit\'{e}, France}
    \author{Laura Hern\'{a}ndez}
    \email{laura.hernandez@cyu.fr}
    \affiliation{Laboratoire de Physique Th\'{e}orique et Mod\'{e}lisation, UMR-8089 CNRS, CY Cergy Paris Universit\'{e}, France}

    \date{\today}

    \begin{abstract}
        In this work we present novel results to the problem of the Hegselmann-Krause dynamics in networks obtained by an extensive study of the behavior of the standard order parameter sensitive to the onset of consensus: the normalized size of the giant cluster. This order parameter reveals the non trivial effect of the network topology on the steady states of the dynamics, overlooked by previous works, which concentrated on the onset of \textit{unanimity}, and allows to detect regions of polarization between the fragmented and the consensus phases. While the previous results on unanimity are confirmed, the consensus threshold shifts in the opposite direction compared to the threshold for unanimity. A detailed finite size scaling analysis shows that, in general, consensus is easier to obtain in networks than in mixed populations. At a difference with previous studies, we show that the network topology is relevant beyond the finitness of the average degree with increasing system size. In particular, in pure random networks (either uniform random graphs or scale free networks), the consensus threshold seems to vanish in the thermodynamic limit. A detailed analysis of the time evolution of the dynamics reveals the role of \textit{bridges} in the network, which allow for the interaction between agents belonging to clusters of very different opinions, after several repeated interaction steps. These bridges are at the origin of the shift of the confidence threshold to lower values in networks as compared to lattices or the mixed population.
        \end{abstract}

    \maketitle

    \section{Introduction}
        The formation and diffusion of opinions in a society, a typical object of study of social sciences, is also of interest for physicists and mathematicians as it can be formulated in terms of complex dynamical systems~\cite{Castellano2009Statistical}. From this perspective, the objective is not to understand the opinion formation in a particular case, but to address questions concerning large scale properties that are observed in many different real systems instead. Several models have been proposed, which aim at understanding how these different observed regularities, like consensus, polarization or fragmentation, emerge from a dynamics governed by the interactions among social actors. These models are stylized simplified versions of real societies which integrate, in general, the most widespread properties of human interactions considered by researchers in social human sciences ~\cite{scott2011social,deutsch1955study}. In the last decade, the generalization of the usage of social networks, allowed for large scale phenomenological studies of the outcomes of these dynamical processes, leading to a wide collection of observations that, though limited to the users of these platforms, provide a widespread view in terms of geographical location, language and cultural aspects of the involved social actors.

        Among the theoretical studies, bounded confidence models~\cite{deffuant2000mixing,hegselmann2002opinion}, address the fact that social actors are rarely influenced by every other one they encounter. Instead, humans have a tendency to be influenced only by those others holding an opinion that is not so different from their own, while ignoring the opinion of those with whom they strongly disagree. This aspect along with the fact that they also have the tendency to conform to the opinion of these close contacts, are considered as basic ingredient of social interaction, known as \textit{homophily} and \textit{social influence}. The most popular bounded confidence models address two different situations: the Deffuant model~\cite{deffuant2000mixing} studies pairwise interactions, where the interacting agents bring their opinions closer after each encounter. In the Hegselmann-Krause (HK) model~\cite{hegselmann2002opinion} on the other hand, the interactions take place in a group, where each agent adopts the average opinion of its own group. The latter has been less investigated, partially due to difficulties related to its algorithmic complexity, which have recently been lifted~\cite{schawe2020open}.

        Nevertheless, several variants of the original HK model, which assumed a mixed population of homogeneous agents, i.e., all of them having the same value of the confidence parameter $\varepsilon$, which determines the neighborhood of interaction of a given agent, have been studied mostly aiming at introducing some heterogeneity in the model. On the one hand, heterogeneity in the agents' properties was considered, addressing the more realistic situation where the confidence that an agent has on the opinions of others is an idiosyncratic property of the agent and not a global characteristic of the society. These works reveal the paradoxical role of heterogeneity in the emergence of consensus. For example, it has been shown that a population mixing two types of agents (open and closed minded ones) \cite{Lorenz2010Heterogeneous} is able to reach consensus even with confidence values below the critical threshold of the homogeneous model. Recently, an extensive study of the fully heterogeneous HK model~\cite{schawe2020open} has established its phase diagram, confirming the particular case mentioned above and revealing the existence of a re-entrant consensus phase in a region of $\varepsilon$ where the homogeneous model leads to fragmentation. Other interesting variants, like models introducing stubborn agents~\cite{HK_stubborn} or introducing the notion of the cost of the opinion change~\cite{schawe_hernandez_cost} have also been studied.

        The other important source of heterogeneity comes from the interaction patterns. In a real society agents do not have the same \textit{a priori} probability of interacting with any other agent, and it is well known that the outcomes of any diffusion process, be it a process of opinion or epidemic spreading, may be different according to the fact that it takes place in networks or in a mixed population \cite{scott2011social,Castellano2009Statistical,newman_book,barrat2008dynamical}. However, few studies of the HK model in networks exist, and given the richness found for the phase diagram of this model in a mixed population, the problem deserves more attention.

        In 2005 S. Fortunato~\cite{fortunato2005consensus} published the first study of the homogeneous version of the HK model in a networked society considering different network topologies. In the author's words: \textit{We claim that $\varepsilon_c $ can take only two possible values, depending on the behaviour of the average degree d of the graph representing the social relationships, when the population $N$ goes to infinity: if d diverges when $N \to \infty$, $\varepsilon_c $ equals the consensus threshold $\varepsilon_i \sim 0.2 $ on the complete graph; if instead d stays finite when $N \to \infty$, $\varepsilon_c = 1/2 $ as for the model of Deffuant et al.}
        This result, which only depends on the average degree and not on the network's topology, seems to imply that consensus is only non trivially reached in a mixed population. In a networked society instead, when $\varepsilon$ is less than $\varepsilon_c = 1/2 $, the almost trivial threshold above
        which unanimity is inevitable for reasonably connected graphs, no consensus is possible.

        These conclusions are drawn from the study of the behavior of the probability that every member of the society carries the same opinion, taken as a measure of consensus. However this variable measures \textit{unanimity}, instead of consensus.
        This situation is very unlikely in a society, where an opinion is considered to be a consensus when it is shared by a large fraction of the population. Using the terminology of the physics of order-disorder transitions, the system is considered to be in a consensus state (the ordered phase) when a giant cluster of social actors holding the same opinion emerges. This doesn't necessarily mean that all the individuals in the society hold exactly the same opinion, but that there exist a group of individuals supporting the dominant opinion whose size is \textit{of the order} of the size of the system, instead. This still allows for the existence of small size clusters with different opinions. Again, using an analogy with physical systems, this is reminiscent of the difference between the onset of the ordered phase and the ground state of the system (perfect order).

        In this work we address the problem of the HK dynamics in networks by an extensive study of the behavior of an order parameter sensitive to the onset of consensus: the normalized size of the giant cluster. This order parameter reveals the non trivial effect of the network topology on the steady states of the dynamics, overlooked by previous works, and allows to detect regions of polarization between the fragmented and the consensus phases. While the previous results on unanimity are confirmed, the consensus threshold shifts in the opposite direction compared to the threshold for unanimity. Detailed finite size scaling analysis shows that, in general, consensus is easier to obtain in networks than in mixed populations. Moreover different network topologies yield different values for the threshold to consensus and in particular, for purely random networks like uniform random graphs or scale free networks, the threshold seems to vanish in the thermodynamic limit.

        The article is organized as follows: in Sec.~\ref{sec:model_methods} we describe the HK model and detail the different lattices and network parameters, as well as the technical calculation details. In Sec.~\ref{sec:results} we present our results grouping them into different categories: Sec.~\ref{sec:results:fully} the fully connected network as a proxy for the mixed population, in Sec.~\ref{sec:results_local} the results concerning pure and perturbed square lattices, where we use the Watts-Strogatz algorithm to introduce shortcuts, and in Sec.~\ref{sec:random} we address the case of purely random models considering two different topologies: the Erd\H{o}s-Renyi graph and the Barab\'asi-Albert network, in Sec.~\ref{sec:dynamics} we analyze the differences in the dynamics corresponding to different topologies. Finally, in Sec.~\ref{sec:discussion} we discuss our results and we present our conclusions and perspectives in Sec.~\ref{sec:conclusions}.

    \section{Models and Methods}
    \label{sec:model_methods}
    \subsection{The Hegselmann-Krause Model}
        The Hegselmann-Krause Model is a model of bounded confidence, where agents may only interact if their opinion differs in less than a \emph{confidence range} $\varepsilon$. Originally it assumes that every agent may interact with any other agent provided that the confidence criterion is fulfilled.
        Here, we study the HK model, additionally constrained by the topology of an underlying static network that materializes the possible social contacts of the agents.
        The update rule changes the opinion of each agent, synchronously, to the average value of its own opinion and those of its neighbors within the confidence range.
        This eventually leads to a final state, in which all agents' opinions do not change anymore.

        Formally, we define the model on a graph $G = (V, E)$, where $V$ is a set of vertices and $E$ a set of undirected edges, representing the agents and their possibility of interaction respectively.
        The set of interaction partners of agent $i$, considering the confidence $\varepsilon$, is therefore
        \begin{align}
            \label{eq:neighbors}
            I(i, \vec{x}) = \left\{ 1 \le j \le n \big| \abs{x_i - x_j} \le \varepsilon \wedge \{i, j\} \in E \right\},
        \end{align}
        where $x_i \in [0,1] $ represents the current opinion of agent $i$.

        The dynamics is defined in discrete time by a synchronous update of all agents at each time step. The opinion of agent $i$ is updated by adopting the average opinion of all the agents in the set given by Eq.~\eqref{eq:neighbors} and its own as:
        \begin{align}
            \label{eq:update}
            x_i(t+1) = \frac{1}{N_i(\vec{x}(t)) + 1} \left[ x_i(t) + \sum_{j\in I(i, \vec{x}(t))} x_j(t) \right]
        \end{align}
        where $N_i(\vec{x}(t))$ is the cardinality of $I(i, \vec{x})$, i.e., the number of  neighbors of the active agent whose opinion at time $t$ is within its confidence range.

        The studied networks are required to be connected simple graphs, i.e., no multi-edges or
        self-loops. As in the original HK model the active agent is explicitly included in the average of Eq.~\eqref{eq:update}. Notice that the latter is required by the synchronous update in order to avoid introducing artificial limit cycles in the dynamics. This can be easily illustrated by considering a system of two agents that are able to interact. Without including the active agent's opinion in the average, the two agents will be trapped in a `flip-flop' cycle, where they swap opinions at each time step. If the active agent's opinion is included in the average, they will assume their average opinion in the first time step.

        In order to avoid finite size effects, it is necessary to perform simulations of very large systems and this is only feasible with a fast algorithm. Here we choose either of two different algorithms depending on the density of the underlying network. For dense networks, i.e.,
        networks, where the average number of neighbors (the \emph{degree}), grows with the number
        of agents $\avg{k} = \mathcal{O}(N)$, we use an efficient tree based algorithm
        introduced in \cite{schawe2020open}. This algorithm offers a quick way to
        identify all agents within the confidence bound which can then be tested to check whether
        they are topological neighbors or not. (Additionally, it offers a fast way to
        calculate the new average opinion at that point of the dynamics when many agents have already converged to the same opinion for the special case of a fully connected network, where the topology check is not necessary).

        On the other hand, for sparse networks, i.e., where the mean degree is independent of the number of
        agents $\avg{k} = \mathcal{O}(1)$, this method becomes less efficient, since the number
        of agents within the confidence bound grows typically like $\mathcal{O}(N)$,
        it is faster to test the neighbors of the active agent (on average $\avg{k}$) with respect to the confidence bound.

        In all cases, we start with the initial opinions $x_i(0), \forall i=1,\ldots,N $
        drawn from a uniform distribution in the range $[0, 1]$. Then we iterate
        Eq.~\eqref{eq:update} until we do not observe changes of the agents'
        opinions anymore, in which case we reached the \emph{final state}. The convergence criterion used here is:
        \begin{align}
            \label{eq:convergence}
            \sum_{i=1}^N \abs{x_i(t-1) - x_i(t)} < 10^{-3}.
        \end{align}
        Note that due to the sum over all agents, this becomes a quite strict criterion for
        large systems, in particular stricter than looking at the opinion change
        of a single agent during a time-step for any threshold above the machine
        precision (we use 32 bit IEEE 754 floating point numbers).

        The main observable of interest is the relative size, $S$, of the largest cluster, defined as
        the set of agents holding the same opinion within a tolerance of $10^{-3}$. In principle an opinion cluster is not necessarily a topological cluster, meaning that agents with the same opinion do not need to be connected in the network. However
        they almost always coincide due to the very strict definition
        of what we classify as the same opinion and the fact that connectivities are not too small.

        Unless stated otherwise, the results shown here are issued from averages over samples of 1000 independent realizations, and we compute the mean relative
        cluster size $\avg{S}$ and its variance $\mathrm{var}(S)$. In order to compare with previous works we also compute the probability of \textit{unanimity}, i.e., the probability that \textit{all} agents have exactly the same opinion (i.e., $S=1$), $P_u$.

        We also study the convergence time $t_c$, i.e.,
        the number of synchronous updates of every agent until the convergence criterion is fulfilled.

        The final states of the system may be classified as:
        \begin{itemize}
            \item \emph{unanimity} if \emph{all} agents have the same opinion, ($\avg{S} = 1$)
            \item \emph{consensus} if one opinion dominates the society, ($\avg{S} \lessapprox 1$)
            \item \emph{polarization} if there are two roughly equally strong opinions, ($\avg{S} \approx 0.5$, coexisting with a second cluster of similar size)
            \item \emph{fragmentation} if there are more than two or no dominating opinions
        \end{itemize}

        It should be noted that while sparse networks use far less computational time to perform one update, we will show later that their dynamics are much slower. As a consequence, the system
        sizes that can be studied for the sparse networks are smaller than for the fully
        connected case. Nevertheless, the sizes studied here which are larger than those considered in previous works, are large enough to extrapolate the tendencies of the system in the thermodynamic limit.

    \subsection{The studied networks}
        We study a selection of network models chosen to cover a large range of topological properties
        going from lattices to random networks, and from fully connected to scale free networks.
        The list considered here is:

    \paragraph{Fully connected network or complete graph}
        Agents arranged in a complete graph can be considered as not being restrained by any topology at all, as each agent is able in principle, to interact with any other agent in the society. This is why it is also referred to as the mixed population society. This corresponds to the setup of the original HK model~\cite{hegselmann2002opinion}. In the thermodynamic limit this is equivalent to a system with infinite interaction range and therefore it is expected to lead to the same results as the mean-field approach.

    \paragraph{Square lattice}
        We use periodic boundary conditions so as to assure perfect regularity in the neighborhood of the agents. In order to vary the connectivity $\avg{k}$, we study lattices that also include connections beyond the nearest neighbors.
        At a difference with other network topologies, lattices are embedded in a $d$-dimensional space. Here we concentrate our study on $d=2$. Lattices show other important differences with respect to random networks: they have a long average path length and a large clustering coefficient.

    \paragraph{Watts Strogatz model or the \emph{small world} network (SW)}
        This network model \cite{watts1998collective} interpolates between lattices and completely random networks. Starting from a one-dimensional lattice with $c$ neighbors, each edge is rewired with a probability $p$ to a randomly chosen node, creating shortcuts in the network. This leads, for small values of $p$, to an ensemble of random graphs with both large clustering coefficient and an average path length that scales as $\avg{l} \propto \log(N)$, known as the \emph{small world} effect.

        Here we build the SW networks starting from the 2-$d$ square lattice, so that the generated SW ensemble acts as a perturbation to the perfect 2-$d$ lattice.

    \paragraph{Barab\'asi Albert (BA)}
        This network model connects vertices using a preferential attachment procedure and leads to a scale free degree distribution in the large $N$ limit~\cite{barabasi1999emergence}.
        This growing network model starts with a clique of $m$ nodes and adds sequentially
        new agents each of which brings $m$ new edges. Each of these edges is connected to the existing core with a probability proportional to the current degree of the target node. In this way nodes with a high degree gain more neighbors in a \emph{`the rich get richer'} procedure. By construction the networks of this ensemble have mean degree
        $\avg{k} = 2m$

    \paragraph{Erd\H{o}s R\'enyi network (ER)}
        Also known as uniform or binomial random graph, any two nodes are connected with probability $p$, leading to a network with uncorrelated connections and a mean degree $\avg{k} = Np$, in the large $N$ limit \cite{erdoes1960}.
        Here we study the sparse version for different fixed values of $Np$. For large $N$ one needs to control that the generated networks are connected in order to avoid misleading results of the HK dynamics.

        Since we are conditioning on connectedness, the generation of ER realizations
        is not entirely trivial. Especially, it is known that almost all ER realizations
        are not connected for a connectivity of $Np < \ln(N)$ in the limit of large $N$ \cite{erdoes1960}.
        In our case, we mainly study $Np = 10$, which prohibits a rejection based sampling
        of connected ER realizations at $N \ge 65536$, the size we have simulated for all
        other ensembles. However, the results for sizes up to $N = 32768$ (which still
        is above the threshold $\ln(32768) \approx 10.4$, but apparently close enough that we
        encounter enough fluctuations of connected ER) already draw a convincing
        picture, such that we do not need to use more sophisticated methods, like Markov chain methods,
        to generate connected ER of larger sizes.

    \section{Results}
        \label{sec:results}
        We will present the results of the HK dynamics organized according to the global characteristics of the assumed social ties.

        Sec.~\ref{sec:results:fully} presents the results for the complete network, which corresponds to the densest topology.
        The order parameter $S$ reveals that the phase transition takes place at $\varepsilon_c \approx 0.1926(5)$, in agreement with well known result the original HK model in a mixed population~\cite{hegselmann2002opinion,Castellano2009Statistical}.

        Sec.~\ref{sec:results_local} presents the results for two-dimensional lattices and its perturbation (SW network) and Sec.~\ref{sec:random} gives the results for purely random networks (ER and BA)

        For any networked society, our results show that the threshold value of the confidence above which consensus can be reached is considerably below $0.2$, which is the threshold for a mixed population.
        This means that if we compare two societies with the same closed minded agents ($\varepsilon < 0.2$) the one with mixed population will not reach consensus while the networked society will.

        Unless stated otherwise, all measured values are averaged over $1000$ realizations per value of
        $\varepsilon$, which are chosen to be $0.001$ apart. To save computational time,
        no simulations above $\varepsilon = 0.6$ are performed, since they trivially
        result in unanimity (almost always). Also we often show curves with $\varepsilon \le 0.3$ to focus
        on the interesting region. The raw data of generated cluster configurations of the final states,
        with the corresponding cluster sizes, opinions and convergence times are available at~\cite{rawData}.

        A summary with best estimates for the transition points $\varepsilon_c$, are given in table~\ref{tab:crit} for easy reference.

        \begin{table}[htb]
            \caption{\label{tab:crit}
                Critical values $\varepsilon_c$ for all ensembles and connectivities $\avg{k}$ we studied.
                The fully connected network shows the transition at $\varepsilon_c = 0.1926(5)$. Note
                that all measurements for BA and ER are compatible with zero within errorbars.
            }
            \centering
            \begin{tabular}{rllll}
                \toprule
                $\avg{k}$ & \multicolumn{1}{c}{$4$} & \multicolumn{1}{c}{$10$} & \multicolumn{1}{c}{$12$} & \multicolumn{1}{c}{$20$}\\
                \midrule
                square lattice    & $0.175(1)$   & --          & $0.0801(7)$ & $0.0548(7)$ \\
                perturbed lattice & --           & --          & $0.0649(8)$ & --          \\
                BA                & $-0.021(53)$ & $0.014(26)$ & --          & --          \\
                ER                & --           & $0.007(50)$ & --          & --          \\
                \bottomrule
            \end{tabular}
        \end{table}

    \subsection{Fully Connected}
    \label{sec:results:fully}
        The main interest of this section, which does not introduce fundamentally new results, is to serve as a reference to compare the effects that appear in networked societies.

        Additionally the results presented here are, to the knowledge of the authors, issued from the finest resolved simulations for system sizes that go one to two orders of magnitude beyond the largest sizes studied before. This is facilitated by an algorithm first introduced in
        Ref.~\cite{schawe2020open}, which is exact within machine precision and enables us to observe finite-size effects and extrapolate to the thermodynamic limit of $N \to \infty$ exceptionally well.

\begin{figure}[htb]
            \centering
            \includegraphics[scale=1]{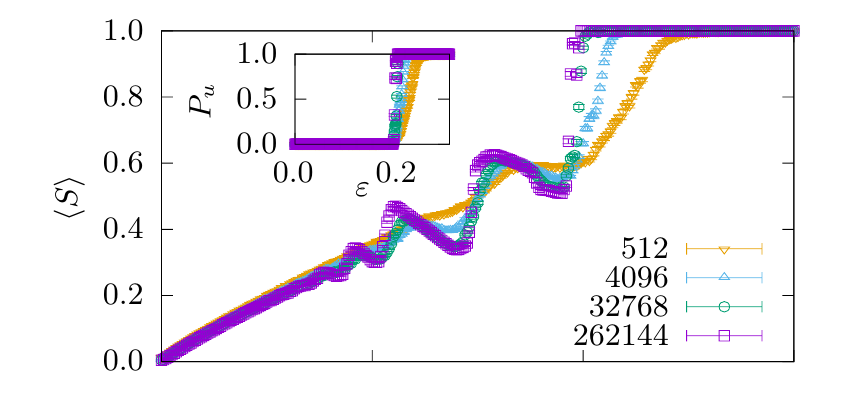}
            \includegraphics[scale=1]{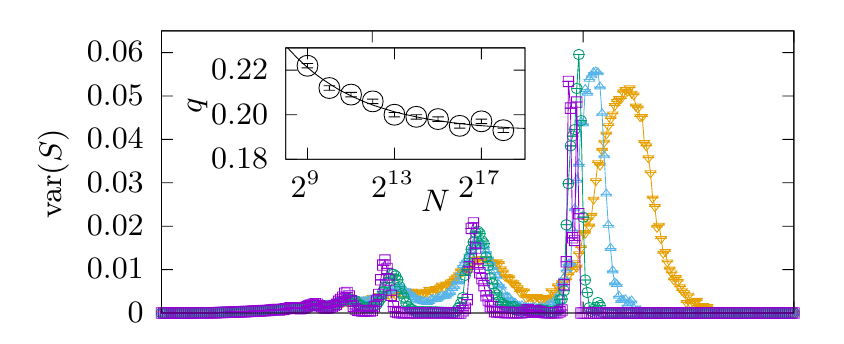}
            \includegraphics[scale=1]{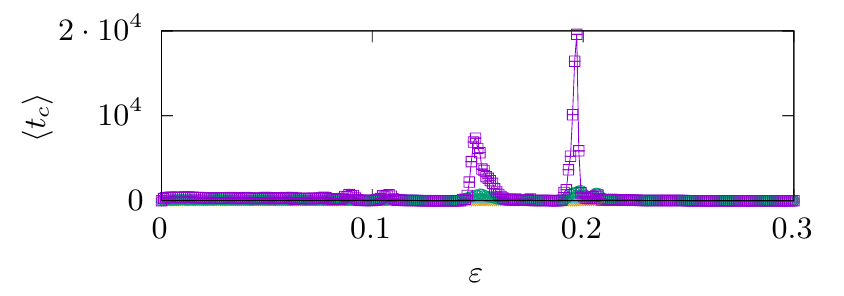}
            \caption{\label{fig:fully}
                Observables of interest for multiple system sizes of the fully connected
                network over 300 values of $\varepsilon \in [0, 0.3]$. Note that all three
                panels share the same horizontal axis. The top panel shows
                the mean size of the largest cluster $\avg{S}$ and the probability of
                unanimity $P_u$ in the inset. The middle panel shows the variance of $S$
                and an extrapolation of position $q$ of its largest peak. The extrapolation
                is a power law with offset $q(N) = \varepsilon_\wedge + aN^{-b}$ (see text for details).
                The bottom panel shows the convergence time $t_c$, each peak of the variance
                coincides with a peak in the convergence time.
                Not all measured sizes are shown for clarity.
            }
        \end{figure}

        Figure~\ref{fig:fully} shows results for systems of sizes up to
        $N = 262144$ agents. The inset of the top panel shows that the transition to unanimity is in agreement with the work in Ref.~\cite{fortunato2005consensus}.

        The introduction of the order parameter allows us to reveal secondary structures that cannot be captured by $P_u$, which indicates the transitions of the society from $n+1$ to $n$
        dominant opinions. Indeed, this \emph{bifurcation} process has already been studied for
        the HK model \cite{fortunato2005vector} and its variants \cite{Lorenz2007continuous,Lorenz2010Heterogeneous}
        as well as for a variant of the Deffuant model \cite{bennaim2003bifurcation}. Also for the Deffuant model it was observed that there are about $\lfloor 1/2\varepsilon \rfloor$ distinct opinion clusters
        in the end~\cite{weisbuch2003interacting}. For the HK model it is known that the bifurcations are dynamical phase transitions \cite{slanina2011dynamical}.

        \begin{figure}[htb]
            \centering
            \includegraphics[scale=1]{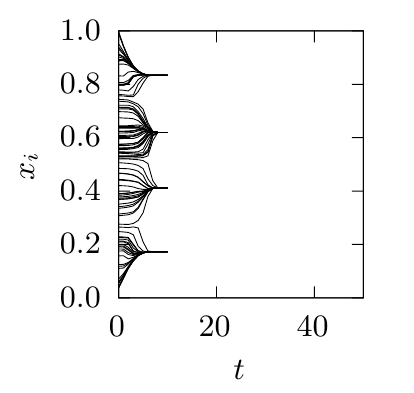}
            \includegraphics[scale=1]{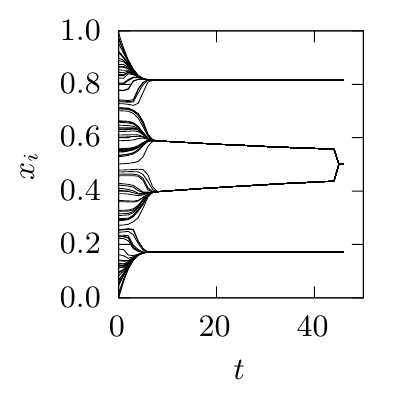}
            \caption{\label{fig:fully_detailed}
                Dynamics causing the steep jumps in $\avg{S}(\varepsilon)$.
                Left $\varepsilon = 0.11$ and right $\varepsilon = 0.12$.
                System of size $N = 16384$ represented by $100$ randomly sampled agents.
                Lines are drawn until convergence.
                The slow merging of the two central clusters in the right panel is facilitated
                by very few agents, which can interact with both clusters
                ($< 1\%$ and therefore not among the randomly sampled agents).
            }
        \end{figure}

        These bifurcations are accompanied by peaks in the variance and in the convergence times,
        as shown in the two other panels of Fig.~\ref{fig:fully}. The peaks in the
        variance (which correspond to a divergence in the susceptibility $\chi = N \mathrm{var}(S)$) are typical
        of continuous phase transitions.
        In order to better understand the large peaks observed in the convergence times we show in Fig.~\ref{fig:fully_detailed} the trajectories in the opinion space before and after a bifurcation
        point. We can see that the bifurcation from 4 to 3 clusters happens in this example through a
        slow merging of the two central clusters, leading to the jump from
        $\avg{S} \approx 1/4$ to $\avg{S} \approx 1/2$, before it relaxes into clusters of equal size
        $\avg{S} \approx 1/3$ (not shown here).
        Indeed, more generally formulated, the mean size of the largest cluster $\avg{S}$ shows that with increasing
        $\varepsilon$ two clusters merge at a $\varepsilon$ value, such that the size of the largest cluster jumps sharply from
        $\avg{S} \approx 1/(n+1)$ to $\avg{S} \approx 2/(n+1)$ and from there it
        decays slower towards $\avg{S} \approx 1/n$ with further increasing $\varepsilon$, before it
        jumps back up at the next `bifurcation', as visible in
        Fig.~\ref{fig:fully}. We can also observe that very large system sizes are needed to
        observe this pattern for $n < 3$.
        Note that the convergence time increases through the merging which is only mediated
        by very few agents---here they are less than $1\%$, such that they are not among the 100 randomly sampled
        agents---and therefore extremely slow, explaining the peaks observed in convergence time plot at the location of the jumps.

        As we will see later, this bifurcation phenomenon does not take place in networked societies, it is characteristic of the mixed population.

        In order to obtain a good estimate of the critical confidence value, $\varepsilon_c$, we will perform finite size scaling analyses both in $\avg{S}$, and in the corresponding variance. This leads to two independent estimates
        of the critical $\varepsilon_c$ backed by the largest simulations to date.

        Currently, the value of this critical threshold is often given as
        $\varepsilon_c \approx 0.2$ \cite{Castellano2009Statistical}. This rough estimate is probably
        issued from two different values obtained for an alternative way to study the HK model, where instead of performing agent based simulations, a density-based HK process is studied. The results depending on the binning chosen for the spatial discretization required to compute the density: $\varepsilon_c \approx 0.22$ \cite{fortunato2005vector}
        and $\varepsilon_c \approx 0.19$ \cite{Lorenz2007continuous}. An even number of
        bins leads to the higher estimate, while an odd number of bins leads
        to the lower one.

        On the contrary, our approach directly simulates the dynamics of the agents such that we do not need an arbitrary discretization (beyond the representation of floating point numbers in a computer). Therefore, since we perform a finite-size analysis going up to very large system sizes, we believe that our estimate for $\varepsilon_c$ is very reliable.

        For continuous phase transitions one can apply \emph{finite-size scaling} close to the critical point in order to collapse the measurements of the order parameter, here $\avg{S}$, (as well as  other magnitudes, like the response functions) onto a size independent scaling form by rescaling the axes. In our case we see that the curves of  $\avg{S}(\varepsilon)$ for all sizes cross in one point, which suggest that they could be  collapsed into a single curve  by rescaling the axes, with  the critical point being located at the crossing point. However here, we observe secondary effects not related to the phase transition to consensus (like the bifurcations), which are interfering with a collapse of the curves. Moreover, this transition does show some signs of a first order transition, like phase coexistence of polarization and consensus states at the transition point instead of the expected scale free distribution of cluster sizes (see Sec.~II of the Supplemental Material \cite{supplemental}), which might be a hint of an hybrid phase transition \cite{Yanqing2011percolation}.
        Nevertheless, with all these precautions,  we  use the position where curves of  $\avg{S}(\varepsilon)$ for different  system size cross as an estimate of the critical point to consensus.
        To do so we estimate the crossing points of each pair of sizes using a linear
        interpolation. These finite-size crossings do not show a clear size dependency, such that we estimate the transition point as the mean of all crossings for large enough sizes
        $N \ge 8192$, which gives: $\varepsilon_\times = 0.1926(5)$.

        We also study how the location of the peaks in the variance curve vary as $N \to \infty $. The position of the  maxima of variance curves $\mathrm{var}(S)$ are an estimate for the steepest slope in $\avg{S}(\varepsilon)$ and should converge to the critical point for large enough sizes. The peak position $q$ is defined as the position of the largest value we measured, i.e., $q = \argmax_\varepsilon \mathrm{var}(S)$. Since the measured values are subject to statistical
        fluctuations, we estimate the uncertainty as twice the spacing between measurements. In order to obtain the peak position in the thermodynamic limit we assume a power law decay towards this asymptotic value, $\varepsilon_\wedge$,
        \begin{align}
            \label{eq:peakfit}
            q(N) = \varepsilon_\wedge + aN^{-b},
        \end{align}
        where $a$ and $b$ are free fit parameters. This power law is motivated from the theory of continuous phase transitions and the fit parameter $b$ is related to the critical exponents of the system.
        The fit is shown in a logarithmic scale for the $N$ axis in the inset of the second panel of Fig.~\ref{fig:fully}. It gives $\varepsilon_\wedge = 0.192(4)$ with a goodness of fit of $\chi^2_\mathrm{red} = 0.57$ and $b=0.42(8)$.
        Both estimates of the critical confidence value, $\varepsilon_c$, show an excellent agreement with the previous mentioned estimates increasing the credibility of both methods.

    \subsection{Lattices: pure and perturbed}
    \label{sec:results_local}
       Lattices are networks that have the particularity of being embedded in a metric space of dimension $d$ and of showing a regular structure.

       In this section we first study the behavior of the HK dynamics on pure lattices, in $d=2$, with interaction ranges that go beyond the nearest neighbors. Then we introduce perturbations to this lattice in the manner of the Watts-Strogatz \textit{small-world} model, so as to study how the results of the HK dynamics change as the interaction network changes from completely regular to random complex networks

    \subsubsection{Square Lattice}
    \label{sec:results:lattice}
        Clearly when studying order-disorder processes in lattices, the dimension where the interactions are embedded is determinant. Here we choose the square lattice, instead of a 1$d$ ring or a 3$d$ structure, because it offers the advantage of an easy visualization, (by coding each agent by a pixel in an image), while avoiding the problem of the lack of phase transition for short range interaction processes in 1$d$.

        Previous results based on the study of $P_u$~\cite{fortunato2005consensus}
        show a sharp transition from fragmentation to unanimity at $\varepsilon_c = 0.5$,
        which suggests that consensus is harder to obtain with limited connectivity. However the order parameter $\avg{S}$ draws a very different picture.

\begin{figure}[htb]
            \centering
            \subfigure[]{\label{fig:lat3:dyn}
                \includegraphics[width=0.45\linewidth]{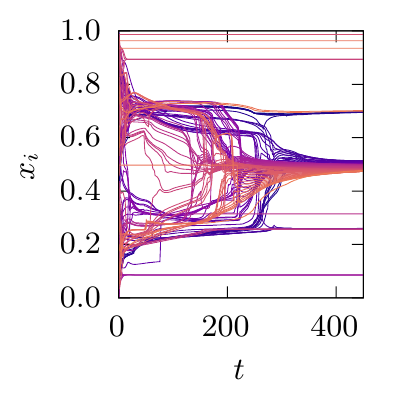}
            }
            \subfigure[]{\label{fig:lat4:dyn}
                \includegraphics[width=0.45\linewidth]{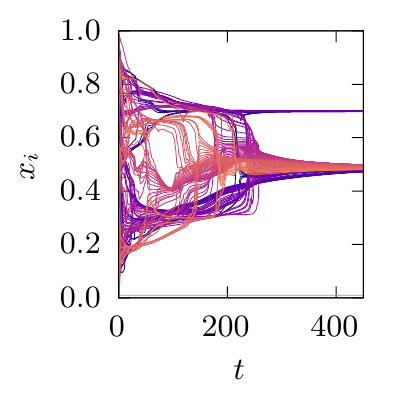}
            }

            \subfigure[]{\label{fig:lat3:pix}
                \includegraphics[width=0.45\linewidth]{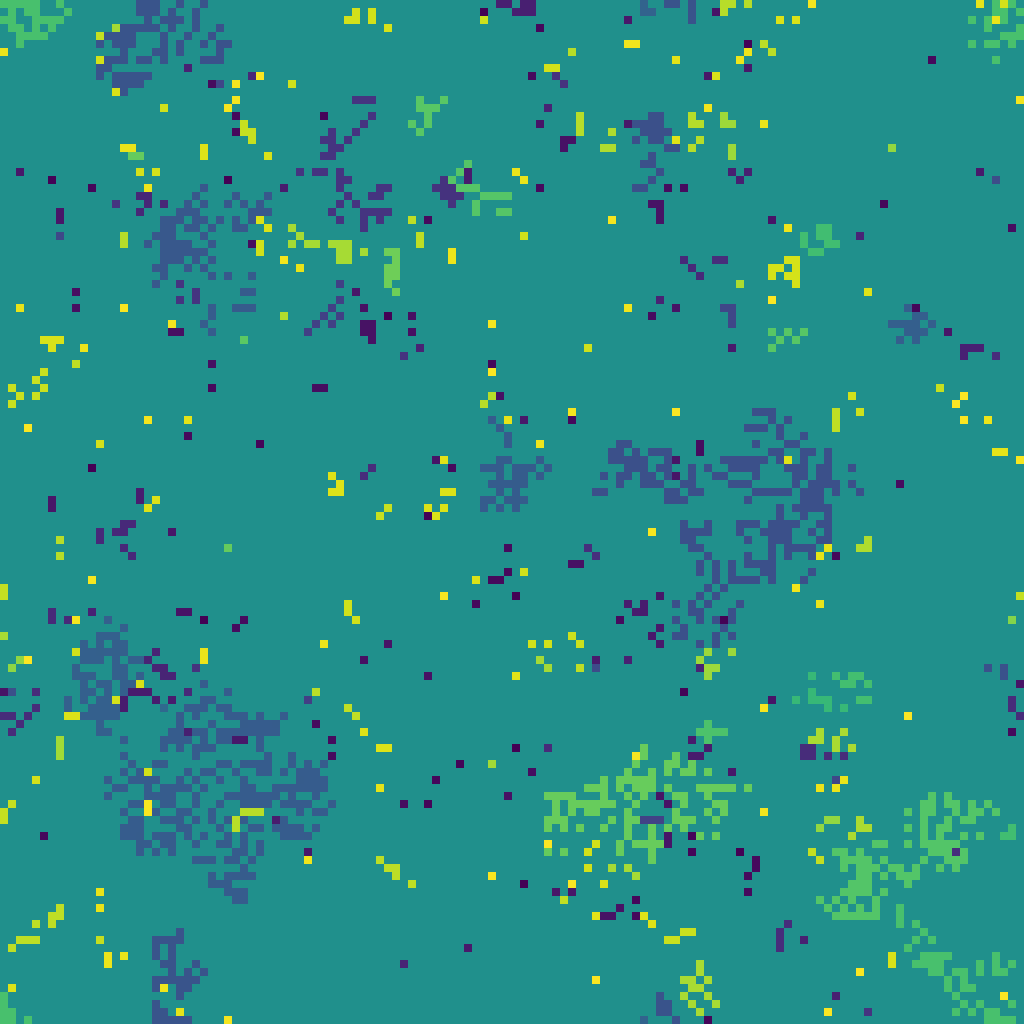}
            }
            \subfigure[]{\label{fig:lat4:pix}
                \includegraphics[width=0.45\linewidth]{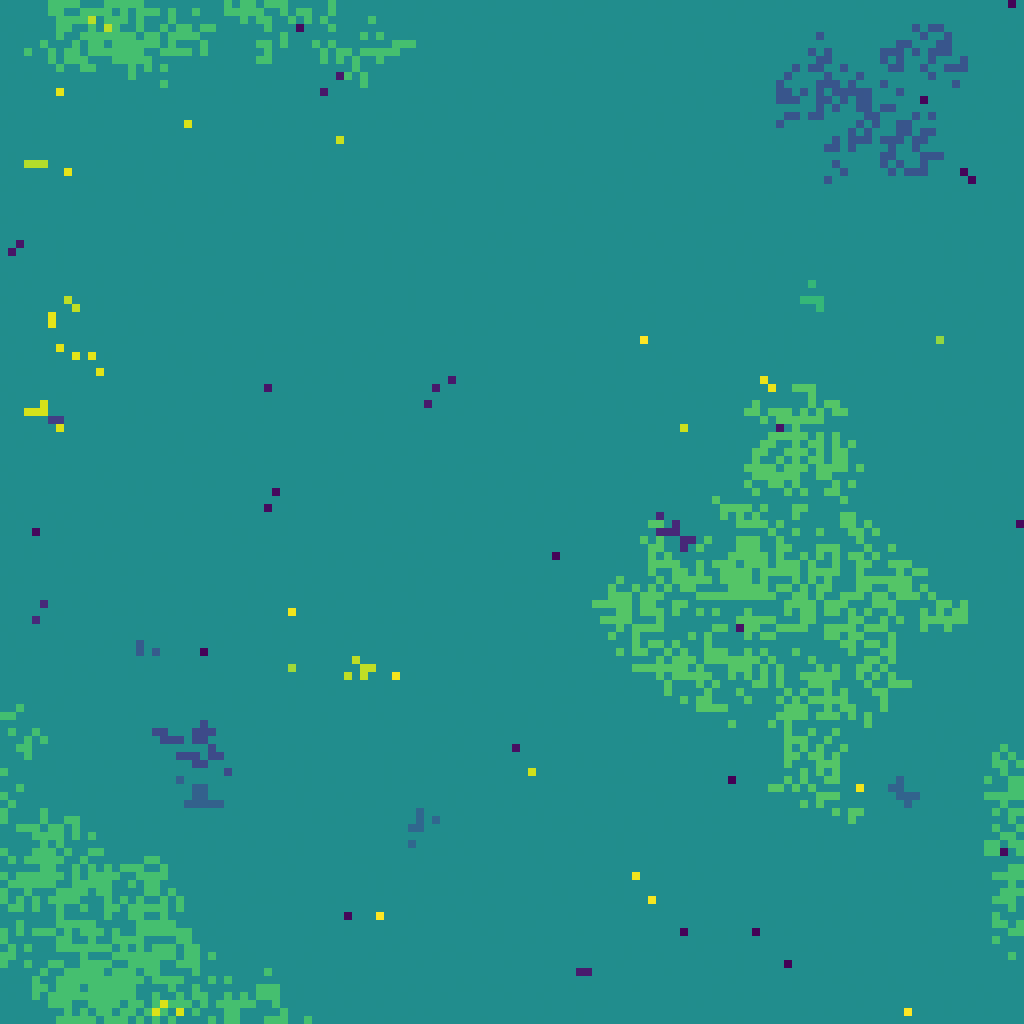}
            }

            \caption{\label{fig:pix_lattice}\label{fig:dyn_lattice}
                Dynamics of \subref{fig:lat3:dyn} a lattice with third nearest neighbors
                and \subref{fig:lat4:dyn} fourth nearest neighbors. The system consists of
                $n = 128^2 = 16384$ agents at $\varepsilon = 0.15$. (The time axis is truncated
                to show the interesting region, convergence according to Eq.~\eqref{eq:convergence}
                is achieved after 4899, respectively, 3465 synchronous updates).
                Both systems end in a final state with one dominating consensus opinion shown in
                \subref{fig:lat3:pix} ($S \approx 0.89$) and \subref{fig:lat4:pix} ($S\approx0.92$).
                The color of the pixels represent the final opinion of the corresponding agent.
                Animations of both evolutions is shown in the Supplemental Material~\cite{supplemental}.
                For clarity not all agents are shown in \subref{fig:lat3:dyn} and \subref{fig:lat4:dyn}
                but only the top line of 128 agents shown in panels (c) and (d) respectively.
                Colors of the lines indicate the starting positions (dark: left, light: right).
            }
        \end{figure}

        Figure~\ref{fig:pix_lattice}, shows the dynamical evolution (top row) and final configurations (bottom row)
        of the opinions of the agents located on the square lattice, with interaction ranging up to the
        third (left column) and to the fourth (right column) neighbors.
        Here, all agents have a confidence of $\varepsilon = 0.15$, well below the critical
        value of the fully connected case. For the sake of clarity on the panels of the top row, only the $128$
        agents of one row of the corresponding lattice are shown, and the colors in the upper panels correspond to a heat map encoding the agent's position of the shown row (going from dark colors at the left border of the network to light at the right one). Considering
        this, we recognize that the convergence towards the central opinion happens for
        groups of agents which are situated nearby. In spite of having different initial opinion values the agents evolve in quite well defined strands towards the central opinion.

        The convergence is much slower than for the mixed population case (note that the scale of the horizontal axis here is an order
        of magnitude larger than that of Fig.~\ref{fig:fully_detailed}).

        The bottom of Fig.~\ref{fig:pix_lattice} shows the final states of the opinion of the
        systems whose dynamical evolution is shown above. Each pixel encodes the final opinion of one agent. A large opinion cluster is clearly observed in both cases, which contains $89\%$ (left) and $92\%$ (right) of all agents.
        Animations of the dynamics of both systems, showing
        very clearly the local character of the spreading of the majoritarian opinion can be found in the Supplemental Material \cite{supplemental}.

        As in a real society, consensus does not imply unanimity, and smaller clusters of different opinions are present in the network.

        The previous discussion shows that consensus is achieved for values of $\varepsilon$ where the mixed population is still in a fragmented state ($\varepsilon_{c}^{\mathrm{mixed}}= 0.192(4)$).

\begin{figure}[htb]
            \centering
            \includegraphics[scale=1]{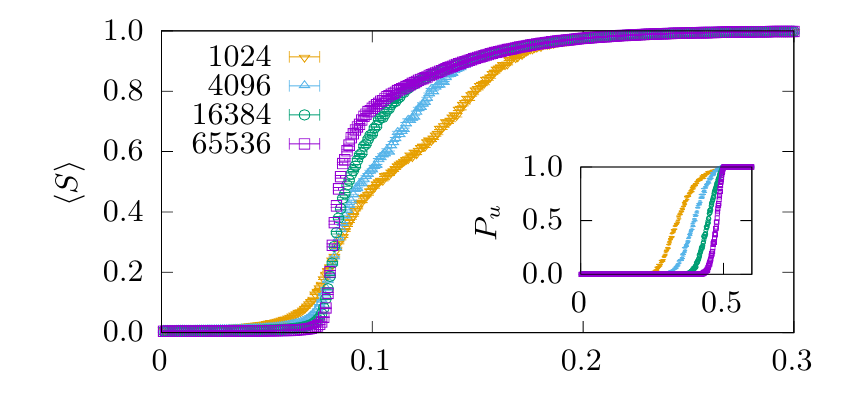}
            \includegraphics[scale=1]{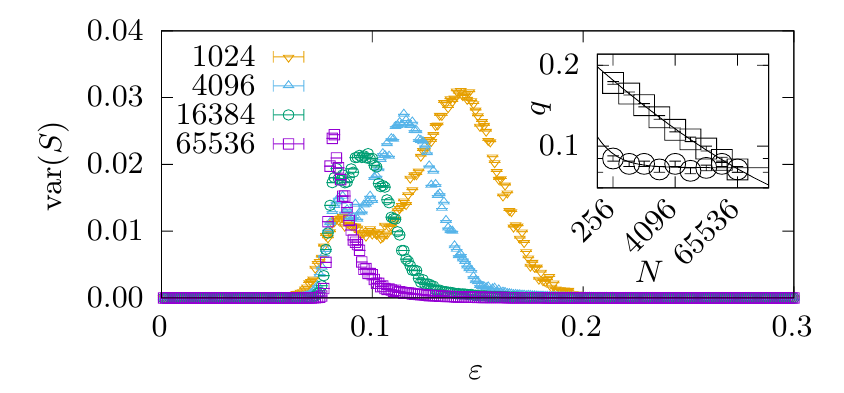}
            \caption{\label{fig:lattice}\label{fig:lattice_var}
                Top panel: Mean size of the largest cluster $\avg{S}$ as a function of the confidence
                $\varepsilon$ for multiple sizes of the square lattice with third nearest neighbors.
                The inset shows the probability of unanimity on a different $\varepsilon$ axis.
                Bottom panel: Variance of the size of the largest cluster $\mathrm{var}(S)$ for the square lattice
                with up to third nearest neighbors. The right peaks converge fast to the
                left (squares in the inset), while the left peaks converge towards the
                transition point (circles in the inset). Note that the left peaks become sharper and higher, as expected for the variance close to a phase transition. The inset shows the positions
                of both peaks in a log-log plot and extrapolations to a power law with offset.
            }
        \end{figure}

        Figure~\ref{fig:lattice} shows the determination of the critical confidence value for lattices. In the top panel the behavior of the order parameter for different system sizes shows that a sharp transition to consensus indeed takes place for values of $\varepsilon < \varepsilon_{c}^{\mathrm{mixed}}$. The inset shows the plot of $P_u$ which leads to the results of Ref.~\cite{fortunato2005consensus}.
        However the main plot shows that the size of the largest
        cluster is actually above $99\%$ already at $\varepsilon = 0.3$ (similar results for nearest and next nearest neighbors may be found in Sec.~III of the Supplemental Material~\cite{supplemental}).
        This figure also shows that finite size effects are important and cannot be neglected. For small sizes, the order parameter seems to indicate the existence of
        two transitions: one from fragmentation to polarization and a second from polarization to consensus, also shown by the two peaks of the variance. However, as $N$ increases the rightmost peaks move to the left while the leftmost peaks stay, merging for the large N limit.

        At a first glance it is clear that the crossing points and therefore the transition towards consensus occurs at a dramatically lower threshold than in the fully connected case.

        In order to determine the critical confidence value, we measure, as before, the crossings of all pairs of sizes
        using a linear interpolation. The transition point is again obtained as the mean over the crossing points
        of all pairs of curves for sizes $N \ge 1024$, which leads to an estimate of: $\varepsilon_\times = 0.0801(7)$.

        The finite size scaling of
        the positions of the variance peaks gives a second estimate of the critical point.
        Interestingly as for the complete network, the variance of the order parameter shows multiple (here two) local maxima. When extrapolating the position of the rightmost peak
        $q = \argmax_\varepsilon \left\{\mathrm{var}(S) \right\}$, it shows a convergence towards $\varepsilon \approx 0$, while the extrapolation the positions of the leftmost peak leads to
        $\varepsilon_\wedge = 0.083(2)$ (with a reduced chi-squared error: of
        $\chi^2_\mathrm{red} = 1.6$), which
        is within two standard errors compatible with the value of the transition point
        obtained from the crossing. This confirms that the rightmost peak is a finite size effect which merges with the left one in the thermodynamic limit. Both extrapolations are shown in the inset of the bottom panel.
        Due to the very low slope and comparatively large relative uncertainties, we can not give any meaningful estimate for $b$.

        Extending the range of interactions to the fourth neighbor, the mean degree almost doubles $\avg{k} = 20$. The results are qualitatively similar to those of the lattice with third neighbor interactions but the impression of a second transition to polarization is enhanced by the existence of a \emph{plateau} in the order parameter and correspondingly, two peaks of its variance. However, again a finite size study of the peaks' position shows that only one remains (more details in Sec.~III of Supplemental Material \cite{supplemental}).

        \begin{figure*}[htb]
            \centering

            \includegraphics[scale=1]{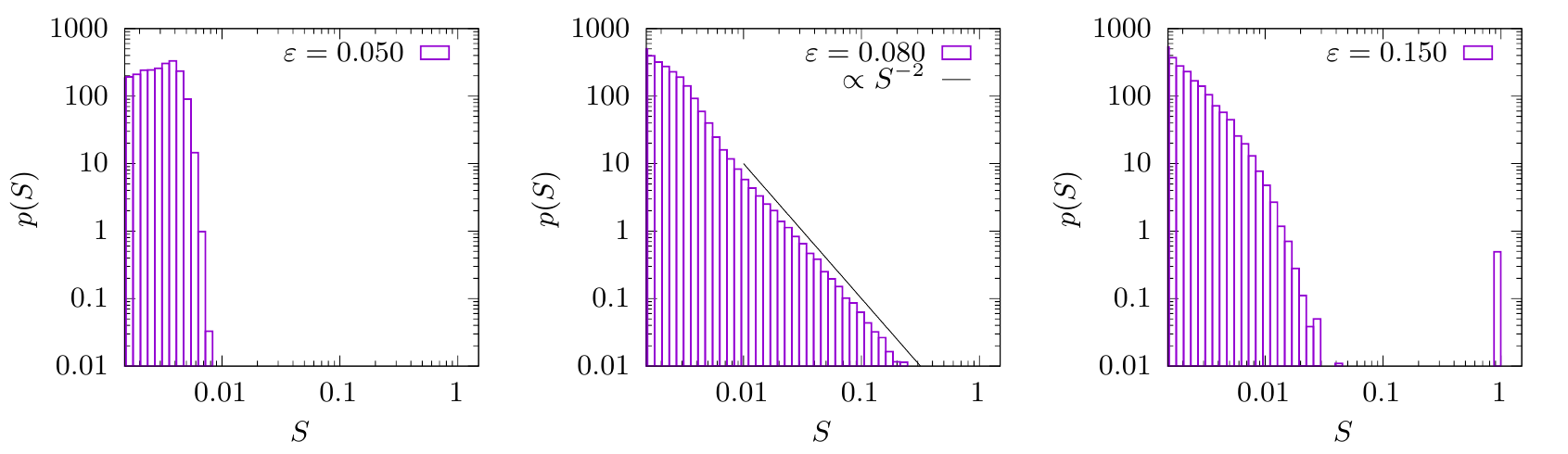}

            \caption{\label{fig:cluster_dist}
                Cluster size distribution $p(S)$ measured in a system with $N=65536$ agents
                averaged over $1000$ realizations for the lattice with up to third nearest neighbors
                below, at ($\varepsilon = 0.080$ is the closest simulated system to our best
                estimate of $\varepsilon_c = 0.0801(7)$) and above the the transition point.
                The bins are logarithmic.
                The line in the central plot is a power law with exponent 2.
            }
        \end{figure*}

        Finally as the transition shows the characteristics of a second order one, it is interesting to study the distribution of cluster sizes, which is expected to be scale free at criticality.
        This means that the probability $p(S)$ to observe a cluster of size $S$ in the system should decay as a power
        law with the \emph{Fisher exponent} $\tau$: $p(S) \propto S^{-\tau}$.

        Figure~\ref{fig:cluster_dist} shows an estimate of the probability density $p(S)$,
        normalized like a continuous distribution, obtained from our data for $N=65536$ using
        logarithmic bins for three values of $\varepsilon$
        below, above and at the critical point.

        As expected, below the threshold (left panel)
        we observe many very small clusters with a fast cut-off, and no cluster spans more than $1\%$ of the system. The right panel shows
        also many small clusters, but also a very pronounced peak close to $S=1$,
        corresponding to the consensus opinion cluster. The probability to observe
        very small cluster sizes is large, simply because a single realization contains many
        very small clusters, but only one large cluster.
        Finally, at the effective
        transition point for this system size (middle panel) the cluster size distribution shows the coexistence of clusters
        of all sizes, according to a characteristic scale-free distribution. Ignoring, as usual, the values
        for very small clusters, we can estimate from the slope in the log-log plot, a Fisher exponent $\tau \approx 2$.

        This observation is another difference with respect to the mixed population case, where only two very sharp peaks appear at $S = 1$ and $S=1/2$ at the transition point (see Sec.~II of the Supplemental Material \cite{supplemental}).

        It has been shown that the entropy of the cluster size distribution is an indicator of the transition point in opinion dynamics models~\cite{opinion_global}, which taking into account the discrete character of the cluster size, can be written as
        % \footnote{Unfortunately $S$ already denotes the relative size of the largest cluster in this manuscript.}
        \begin{align}
            \sigma = - \sum_{z \in Z} p_z \ln(p_z),
        \end{align}
        where $p_z$ is the discrete probability for an arbitrary agent to be member of a cluster of relative size $z$,
        i.e., to contain $zN$ agents, normalized to $1 = \sum_{i \in Z} p_z$, and $Z$ is the set of all occurring relative cluster sizes.

\begin{figure}[htb]
            \centering

            \includegraphics[scale=1]{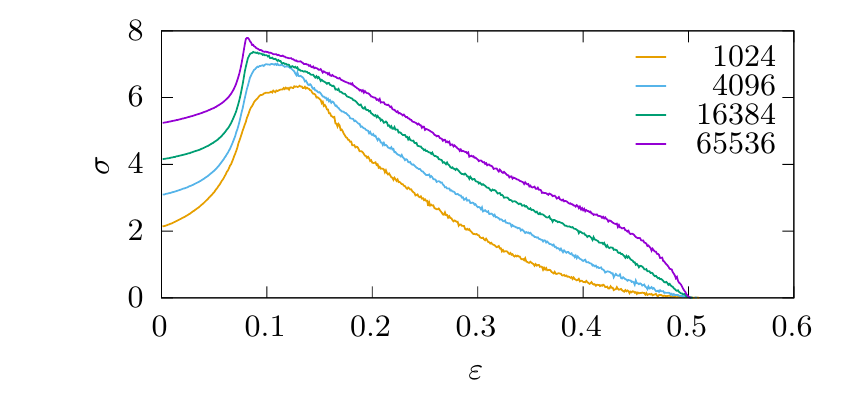}

            \caption{\label{fig:lat3_entropy}
                Entropy $\sigma$ of the cluster-size distribution for the square lattice with up to
                third nearest neighbors. The entropy peaks, sharper for larger sizes,
                at the transition point $\varepsilon_c = 0.0801(7)$. Also, beyond the
                threshold of unanimity $\varepsilon_u = 0.5$ the entropy is 0, since
                beyond this point, every single agent has the same opinion, i.e.,
                the distribution becomes a $\delta$ peak.
            }
        \end{figure}

        Notice that the cluster size distributions
        of unanimity has an entropy of $\sigma = 0$.
        At a phase transition the entropy is expected to show a peak, corresponding to the wide
        power law distribution of cluster sizes. This can be seen in Fig.~\ref{fig:lat3_entropy}.

    \subsubsection{Perturbing the Lattice: Watts-Strogatz network}
    \label{sec:results:ws}

        In order to introduce some randomness in the network of possible social interactions while conserving some properties of the lattices, we slightly perturb the square lattice using
        the rewiring technique introduced by Watts and Strogatz \cite{watts1998collective}
        to study the small world effect.

        Starting from square lattices with interactions up to third neighbors ($k=12$), the algorithm visits each of the edges and with probability $p=0.01$
        disconnects one of its end-points and rewires it to a randomly chosen node, avoiding self-loops and multi-edges. We check that all generated
        instances are connected. For this value of $p$, the obtained network has mostly local regular interactions along with some long range ones.

        The behavior of $\avg{S}(\varepsilon)$ looks very similar
        to the pure lattice, with however, a more pronounced plateau of polarization which does not vanish for the system sizes studied. One example of this behavior is plotted later in the summarizing Fig.~\ref{fig:cmp} and the details can be found in Sec.~IV of the Supplemental Material \cite{supplemental}.

        Again we can obtain by finite size analysis a crossing point leading to a critical value of the confidence
         $\varepsilon_c =0.0649(8)$. This value is clearly lower than for
        the pure lattice, suggesting that random edges facilitate consensus at even lower
        values of $\varepsilon$. This phenomenon leads us to a detailed study of graph ensembles, containing networks with only random connections.

    \subsection{Ensembles of random networks }
    \label{sec:random}
       In this section we discuss the results of the HK dynamics on networks having two very different
        topologies: the Barab\'asi Albert (BA)
        and Erd\H{o}s R\'enyi (ER) random graph ensembles. Our results strongly suggest that the transition point to consensus shifts to $\varepsilon_c \to 0$ in the $N\to \infty$ limit. This means that, unlike the case of lattices and the mixed population, and as opposed to well known previous results~\cite{fortunato2005consensus, Castellano2009Statistical} systems of infinite size will always reach consensus independently of the level of closed mindedness of its agents. The mechanism of this apparently paradoxical result is analyzed in the following sections.

    \subsubsection{Barab\'asi Albert networks}
    \label{sec:results:sf}
\begin{figure}[htb]
            \centering
            \includegraphics[scale=1]{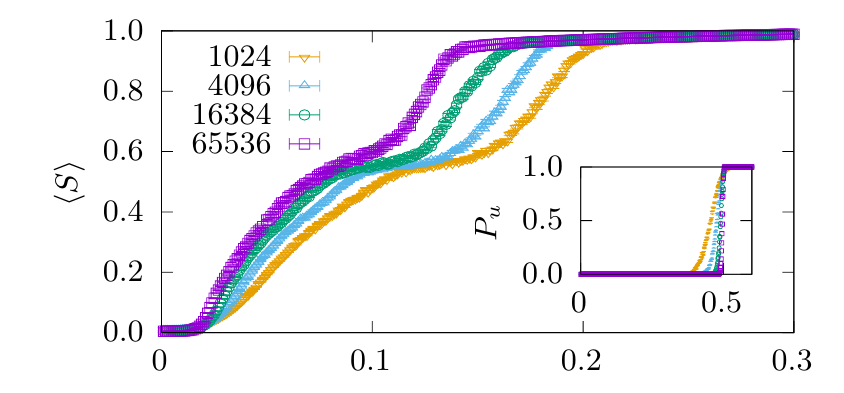}
            \includegraphics[scale=1]{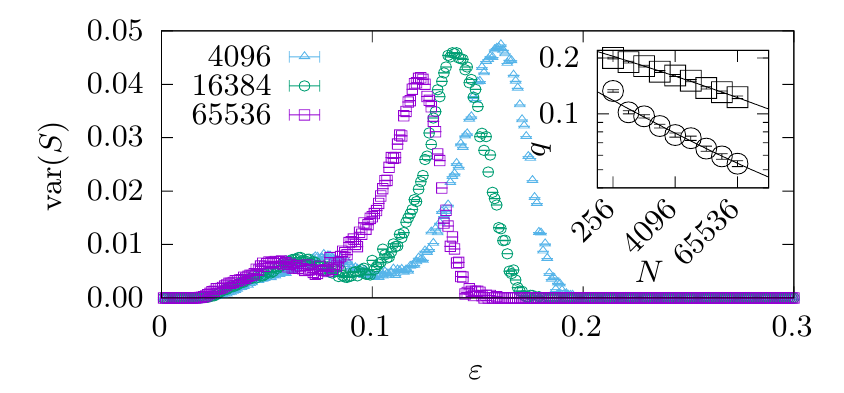}
            \caption{\label{fig:ba}
                Top panel: Mean size of the largest cluster $\avg{S}$ as a function of the confidence
                $\varepsilon$ for multiple sizes of the Barab\'asi Albert ensemble with a mean degree
                of $\avg{k} = 10$.
                The inset shows the probability of unanimity on a different $\varepsilon$ axis.
                Bottom panel: Variance of the size of the largest cluster $\mathrm{var}(S)$ for the
                same ensemble. The right peaks converge as a power law to $0$ (squares in the inset),
                while the left peaks converge even faster towards $0$ (circles in the inset).
                The inset shows the positions of both peaks in a log-log plot and extrapolations
                to a power law with offset. Fit estimates of the offset are always 0 within
                statistical uncertainty; therefore the lines look straight in log-log scale.
                % left peak extrapolation: offset = $0.014(25)$, $\chi = 0.7$
            }
        \end{figure}

        This network topology is characterized by the important fact that its degree distribution becomes scale free in the large $N$ limit.
        Figure~\ref{fig:ba} shows the
        order parameter $\avg{S}$ and its variance for different values of $\varepsilon$.
        First, we notice the curves of $\avg{S}$ for different sizes do not seem to cross at any finite value of $\varepsilon$. Therefore, to determine the critical
        point we study the evolution of the position of the
        variance's peak position. The inset of the bottom panel of Fig.~\ref{fig:ba} shows
        both peak positions, $q$, in a log-log plot for different sizes. At a difference with the results obtained for lattices,
        both positions apparently follow power laws converging to $\varepsilon_\wedge = 0$.
        The value obtained by a fit to the form \eqref{eq:peakfit} is indeed compatible with 0 and
        listed in table \ref{tab:crit}.
        The left peak has even a steeper slope than the right peak, such that they
        never meet and the polarization region persists even for large sizes (over a very small range).

\begin{figure}[htb]
            \centering
            \includegraphics[scale=1]{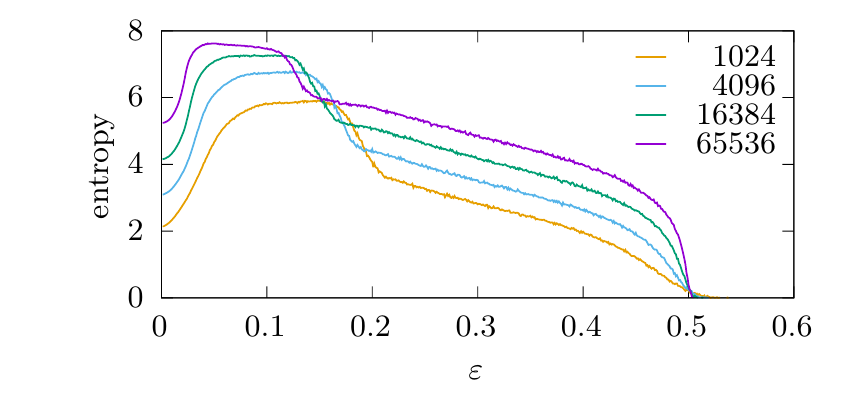}
            \caption{\label{fig:ba_ent}
                Entropy $\sigma$ of the cluster-size distribution for the Barab\'asi Albert
                ensemble with a mean degree of $\avg{k} = 10$. It does not exhibit a sharp peak at
                finite values of $\varepsilon$. Its maximum is stretched over a plateau, which
                moves continuously towards $0$, which is consistent with the before estimated
                transition point at $0$. Also, beyond the
                threshold of unanimity $\varepsilon_u = 0.5$ the entropy is 0, since
                beyond this point, every single agent has the same opinion, i.e.,
                the distribution becomes a $\delta$ peak.
            }
        \end{figure}

        The behavior of the entropy shown in Fig.~\ref{fig:ba_ent}, confirms this observation: there is no sharp peak at any $\varepsilon$, and the width of the plateau located between the $\varepsilon$ values corresponding to two peaks of the variance, diminishes with increasing system size.

        Note that for even sparser networks, e.g., $\avg{k} = 4$, we observe only one peak in the variance.
        Extrapolating this single peak again leads to a limiting transition point of $\varepsilon_c = 0$ (see Sec.~V of the Supplemental Material \cite{supplemental} for more details.)

\begin{figure}[htb]
            \centering
            \includegraphics[width=\linewidth]{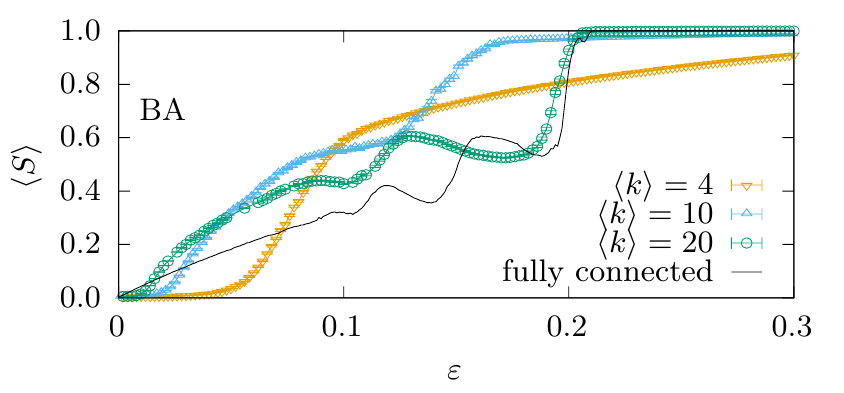}
            \caption{\label{fig:ba_K}
                Development of BA with increasing degree. As expected higher mean degrees $\avg{k}$
                become more similar to the fully connected case (black line). $N=16384$, $1000$ realizations
                per data point, resolution of $0.002$.
            }
        \end{figure}

        To conclude the treatment of the BA ensemble, we investigate the influence of the
        connectivity on the behavior of the order parameter. In Fig.~\ref{fig:ba_K} we plot $\avg{S}(\varepsilon)$ for
        a selection of $\avg{k}$ values, for a system of size $N=16384$.

        As expected, for fixed $N$, higher connectivity rapidly leads to a behavior similar to the one observed for the fully connected case. In particular, the shape of $\avg{S}(\varepsilon)$ changes qualitatively: for very sparse networks only one flank for the increase of $\avg{S}$ is observed, but more appear as $\avg{k}$ increases recalling the bifurcation effect observed in the fully connected case.

    \subsubsection{Erd\H{o}s R\'enyi networks}
    \label{sec:results:er}
        Here we present the results of the HK dynamics constrained by a network with a homogeneous, Poissonian degree distribution.

        Our simulations are able to show that the HK model behaves qualitatively similar on the ER and on the BA ensemble. Interestingly, this means that it is not the scale free nature of the BA network that is responsible for the tendency $\varepsilon_c \to 0 $ in the thermodynamic limit.

\begin{figure}[htb]
            \centering
            \includegraphics[width=\linewidth]{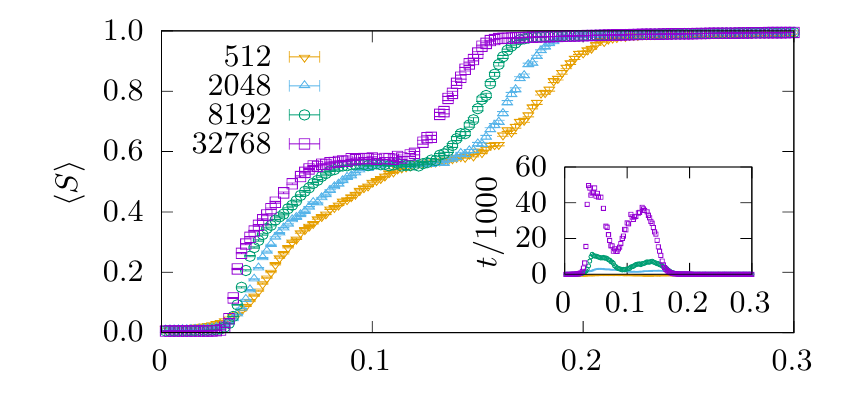}
            \includegraphics[width=\linewidth]{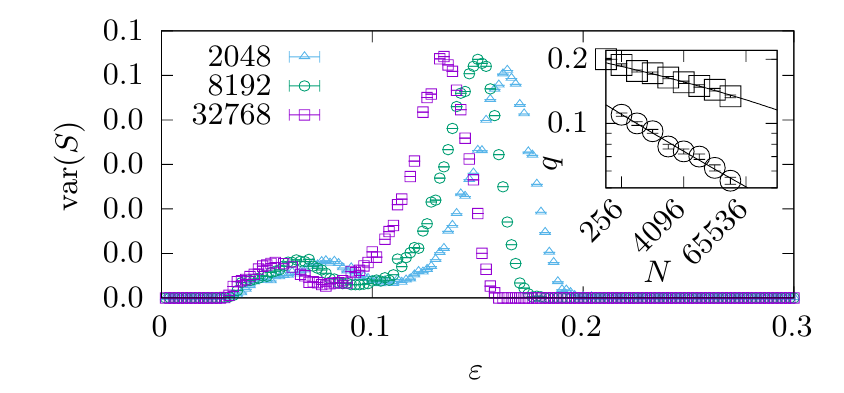}
            \caption{\label{fig:er}
                Top panel: Mean size of the largest cluster $\avg{S}$ as a function of the confidence
                $\varepsilon$ for multiple sizes of the Erd\H{o}s R\'enyi ensemble with a mean degree
                of $\avg{k} = 10$.
                The inset shows the convergence time.
                Bottom panel: Variance of the size of the largest cluster $\mathrm{var}(S)$ for the
                same ensemble. The right peaks converge as a power law to $0$ (squares in the inset),
                while the left peaks converge even faster towards $0$ (circles in the inset).
                The inset shows the positions of both peaks in a log-log plot and extrapolations
                to a power law with offset. Fit estimates of the offset are always 0 within
                statistical uncertainty; therefore the lines look straight in log-log scale.
            }
        \end{figure}

        In the first panel of Fig.~\ref{fig:er}, one can observe the same pattern in the order parameter that we found for the HK in the BA network, although with a more pronounced plateau. The inset shows that the convergence times curve diverges at both flanks of the $\avg{S}(\varepsilon).$ %(note however that the exact values depend somewhat on the abort criterion used).

        The bottom panel shows the variance of the order parameter, also similar to the BA case. In the inset one can see that the extrapolations of the left and right peak positions
        tend both to zero as power laws. The behavior of the entropy of the cluster distribution sizes is also very similar to the one observed for the BA network and is shown in Sec.~VI of the Supplemental Material \cite{supplemental}.

        All these observations show that the shift of the critical point $\varepsilon_c \to 0$ in the thermodynamic limit, is a characteristic of purely random networks, independent of the properties of the degree distribution, which is absent from the HK model on lattices, either pure or perturbed.

\begin{figure}[htb]
            \centering
            \includegraphics[width=\linewidth]{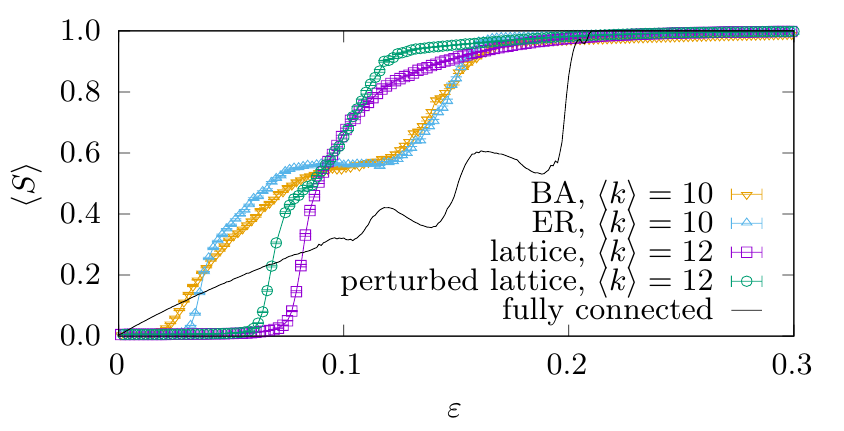}
            \caption{\label{fig:cmp}
                Comparison of all studied ensembles. Regular structures ((perturbed) lattice) inhibit any
                kind of cluster sizes at extremely low values of $\varepsilon$. Non-local structures caused
                by, e.g., random edges (ER, BA, perturbed lattice, and fully connected) lead to a more
                pronounced plateau. $N=16384$, $1000$ realizations per data point, resolution of $0.002$.
            }
        \end{figure}

        Finally, we show in Fig.~\ref{fig:cmp} the comparison of the order parameter $\avg{S}(\varepsilon)$ curves for all the topologies studied here at the same size and for comparable average degree, where two qualitatively different behaviors can be observed. On the one hand completely random networks (BA and ER ensembles) show a similar behavior of $\avg{S}(\varepsilon)$, including a wide region of polarization (plateau), in spite of the fact that their topologies are rather different. On the other hand, the behavior of $\avg{S}(\varepsilon)$ for lattices is completely different: while there is a much less pronounced plateau for pure lattices (barely visible for these parameters), it appears as some random perturbation is included.

        These two groups of interaction networks differ by two main characteristics: lattices are by
        definition embedded in a metric space (here a square lattice) and have a large clustering coefficient. In contrast, ER and BA ensembles are not spatial networks which show, in comparison, low clustering, and may present \textit{bridges}, constituted by a chain of several nodes that join two communities in the network. We will see later how the latter are responsible of the shift to the very low critical confidence value ($\varepsilon_c \to 0$) in the thermodynamic limit.

        From Fig.~\ref{fig:cmp} it could be interpreted that lattices allow consensus to be formed at lower confidences than random networks, however this is a size effect as we have seen that the rightmost flank of the curves corresponding to random networks shifts left very fast with increasing system size.

    \section{The details of the different dynamics}
        \label{sec:dynamics}
        We can roughly summarize our results by saying that on one hand, at a difference with previous results ~\cite{Fortunato2004Krause}, consensus is easier to reach in any networked society than in the mixed population. On the other hand, among networked societies we observe that random networks, will always reach consensus in the thermodynamic limit although this involves diverging convergence times, while there is a finite (low) critical confidence value $\varepsilon$ for the setting up of consensus in lattices (either pure or perturbed).

        In order to understand these apparently paradoxical results, it is necessary to examine in detail the dynamical processes that occur in these different cases.

        The fundamental mechanism responsible for this effect is closely related to the slower dynamics that takes place on all networked societies with respect to the mixed population. As in the latter, agents are not constrained by different fixed neighborhoods and can interact with any other agent within their confidence interval, they very quickly adopt the average opinion of all agents in their opinion neighborhoods. Typically, at low $\varepsilon$, agents on the borders of the opinion space ($x < \varepsilon$ or $x > 1 - \varepsilon$) move to more central opinions, increasing the density of agents near the boundaries of the opinion space. As a consequence, more central agents are attracted to these dense groups, near the boundaries of the opinion space. This effect quickly leads to polarization for intermediate $\varepsilon$ values, or to fragmentation into multiple clusters, for very low values of $\varepsilon$. For large $\varepsilon$ values, agents at the borders are able to move directly to the central opinion and consensus is easily achieved.

        It is worthwhile noticing that the increase of convergence times is observed close to $\varepsilon_c$ due to the slow merging of already collapsed clusters
        mediated by very few agents which can
        interact with both, given that their opinions are within the confidence values of both clusters (cf.~Fig.~\ref{fig:fully_detailed}).

        In networked societies an agent cannot interact with all other members of the society who carry an opinion within its confidence interval. Instead each agent must find in its topological neighborhood other agents within its confidence interval.

\begin{figure}[htb]
            \centering
            \includegraphics[scale=0.75]{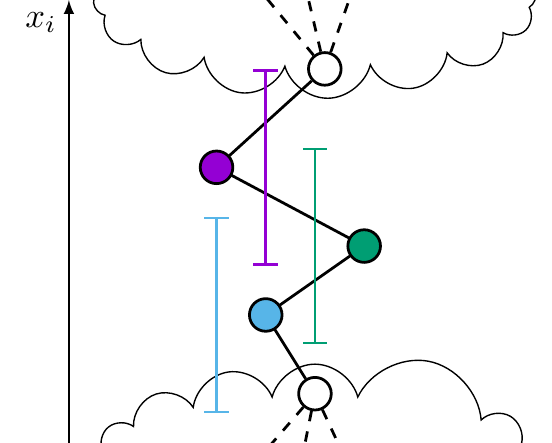}
            \includegraphics[scale=1]{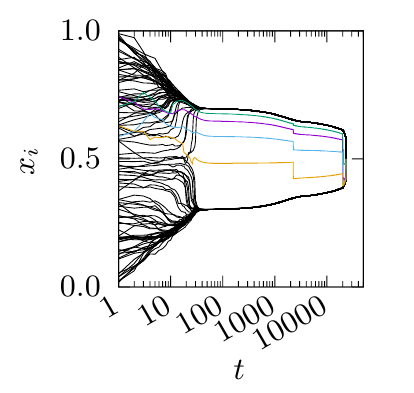}
            \caption{\label{fig:ba_detailed}
                The left panel shows a sketch of a bridge. The line segments show the confidence range
                $[x_i - \varepsilon, x_i + \varepsilon]$ of the corresponding agent marked in the same color.
                The colored nodes will not merge into
                the large clusters (at the top and bottom of the sketch), since the interaction is mediated
                by a similar amount (here the same) of active edges (according to Eq.~\eqref{eq:neighbors})
                to the large clusters as to other members of the chain, such that they enable both large
                clusters to approach each other. The horizontal axis is arbitrary.
                The right panel shows an evolution of a BA realization with $N=16384$ agents at $\varepsilon = 0.14$.
                Shown are 100 randomly selected agents in black on a logarithmic horizontal axis to
                resolve details at all important time scales: First polarization is reached
                on a fast timescale, then the clusters merge on a slow timescale using bridges
                of mediating agents, which are so rare, that none were selected in the random samples of 100
                agents. Therefore some of the mediating agents are shown as colored lines.
            }
        \end{figure}

        In random networks, different opinion clusters may be topologically connected by \emph{bridges} of several agents holding opinions that are different among themselves and also different to the two clusters. This may help to merge clusters whose opinion differ in more than $2 \varepsilon$.
        This is illustrated in the scheme of the left panel of Fig.~\ref{fig:ba_detailed}.

        In sparse networks, agents at the ends of a bridge usually have one or very few neighbors inside the nearest large cluster, such that the influence of the cluster on its own opinion has the same weight as that of another agent of the bridge, and can therefore play its mediating role indefinitely. Such bridges are rare configurations, but very few of them are enough to facilitate consensus. As their probability increases with the network size, we observe that the transition point shifts to arbitrarily low values of $\varepsilon$.

        \begin{figure*}[htb]
            \centering
            \includegraphics[width=0.24\linewidth]{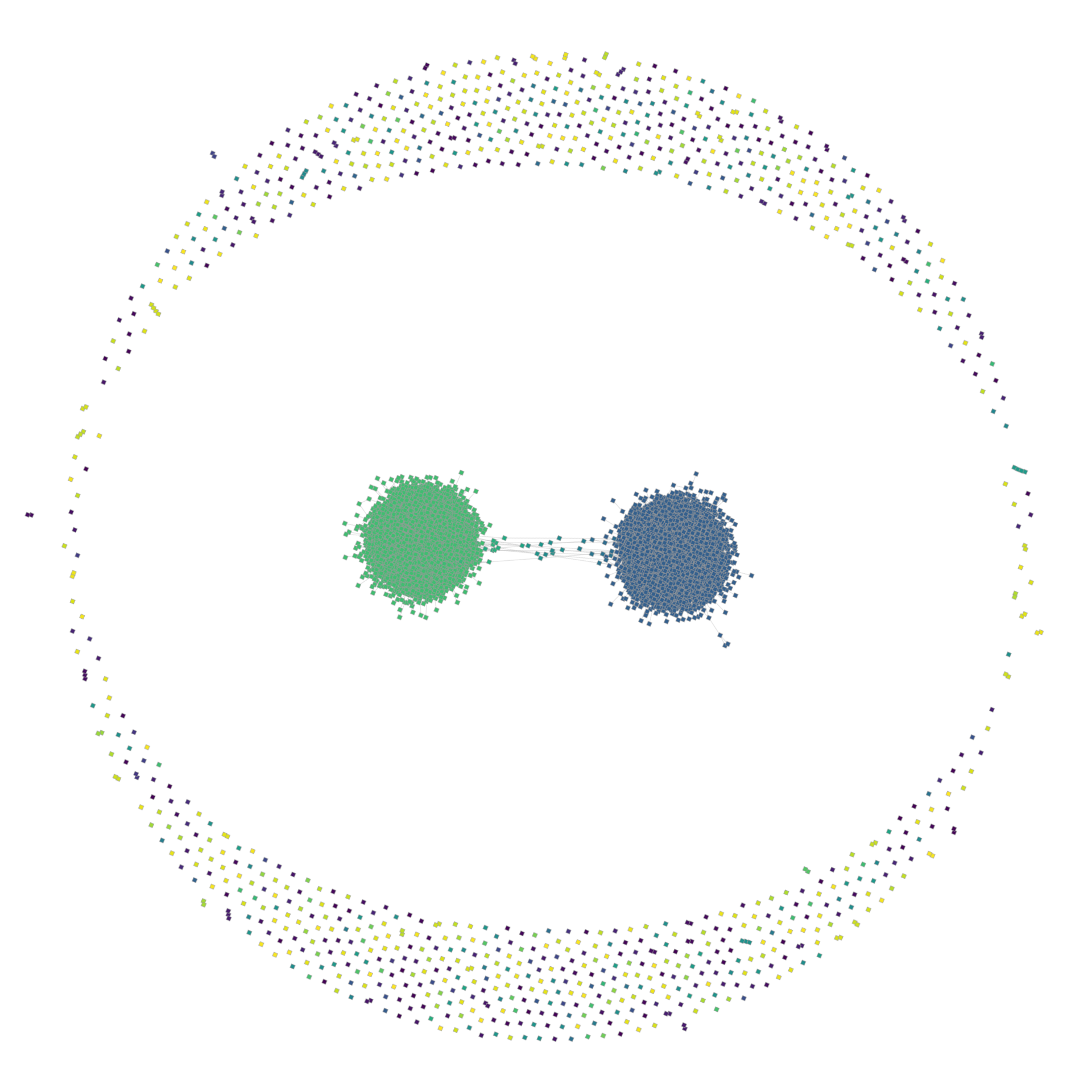}
            \includegraphics[width=0.24\linewidth]{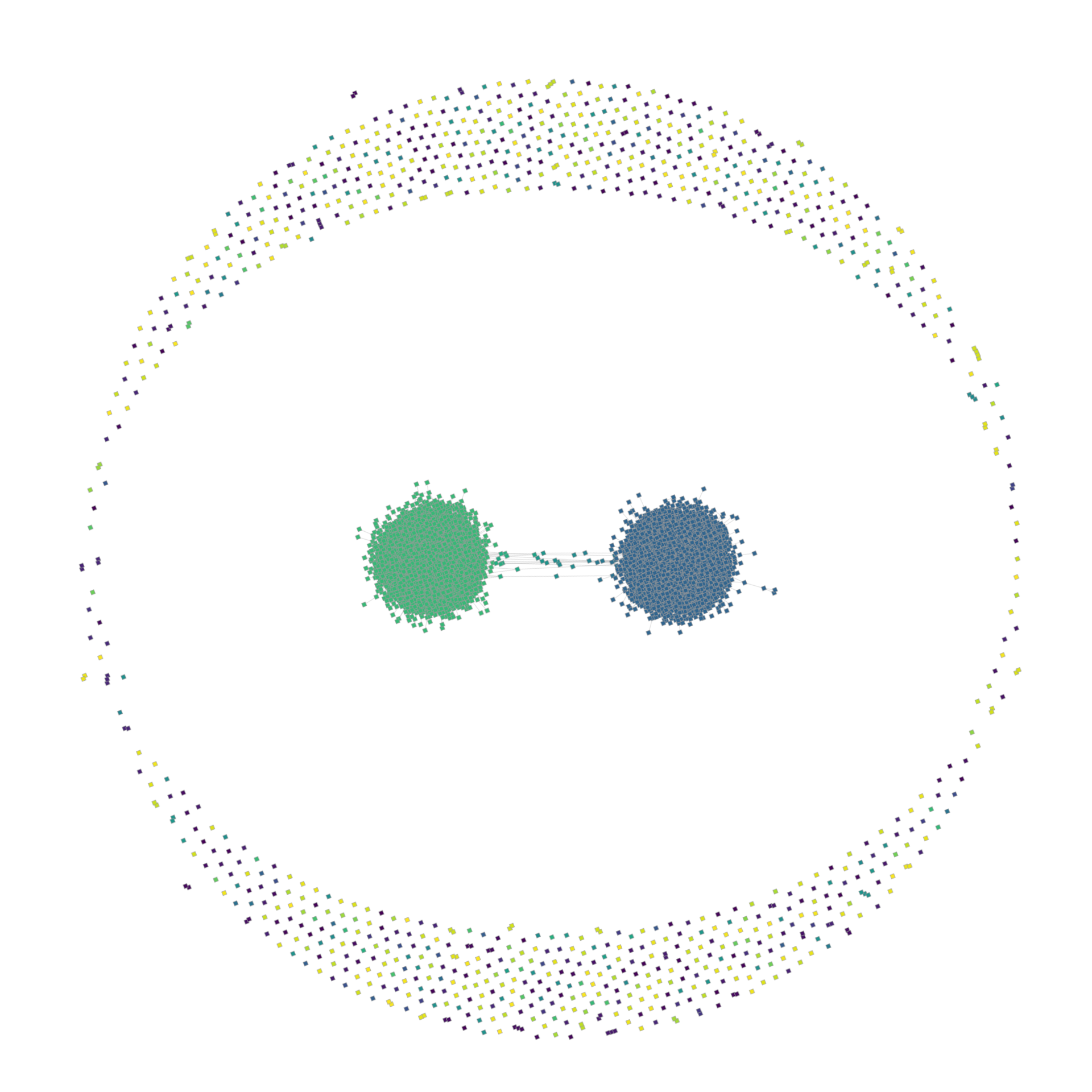}
            \includegraphics[width=0.24\linewidth]{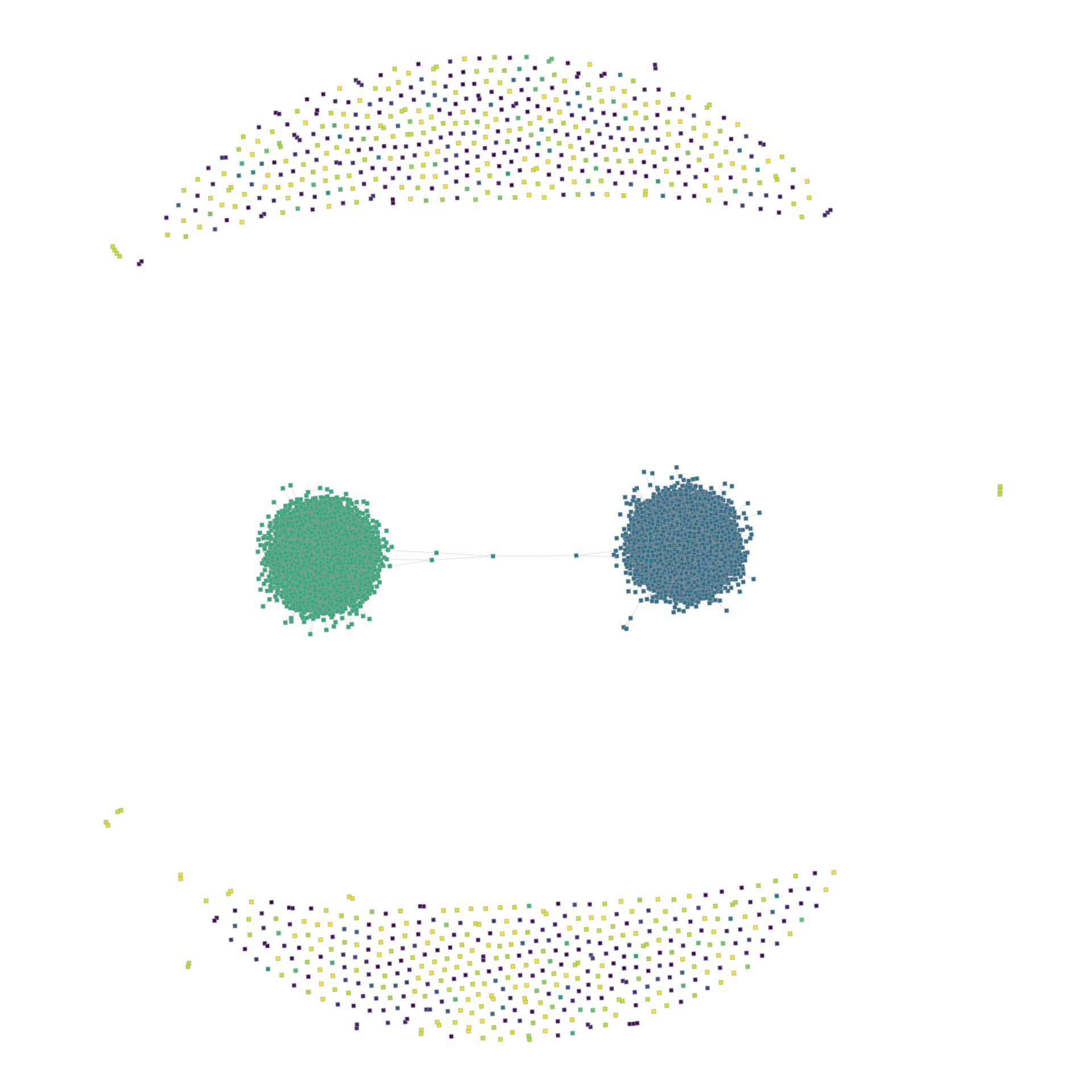}
            \includegraphics[width=0.24\linewidth]{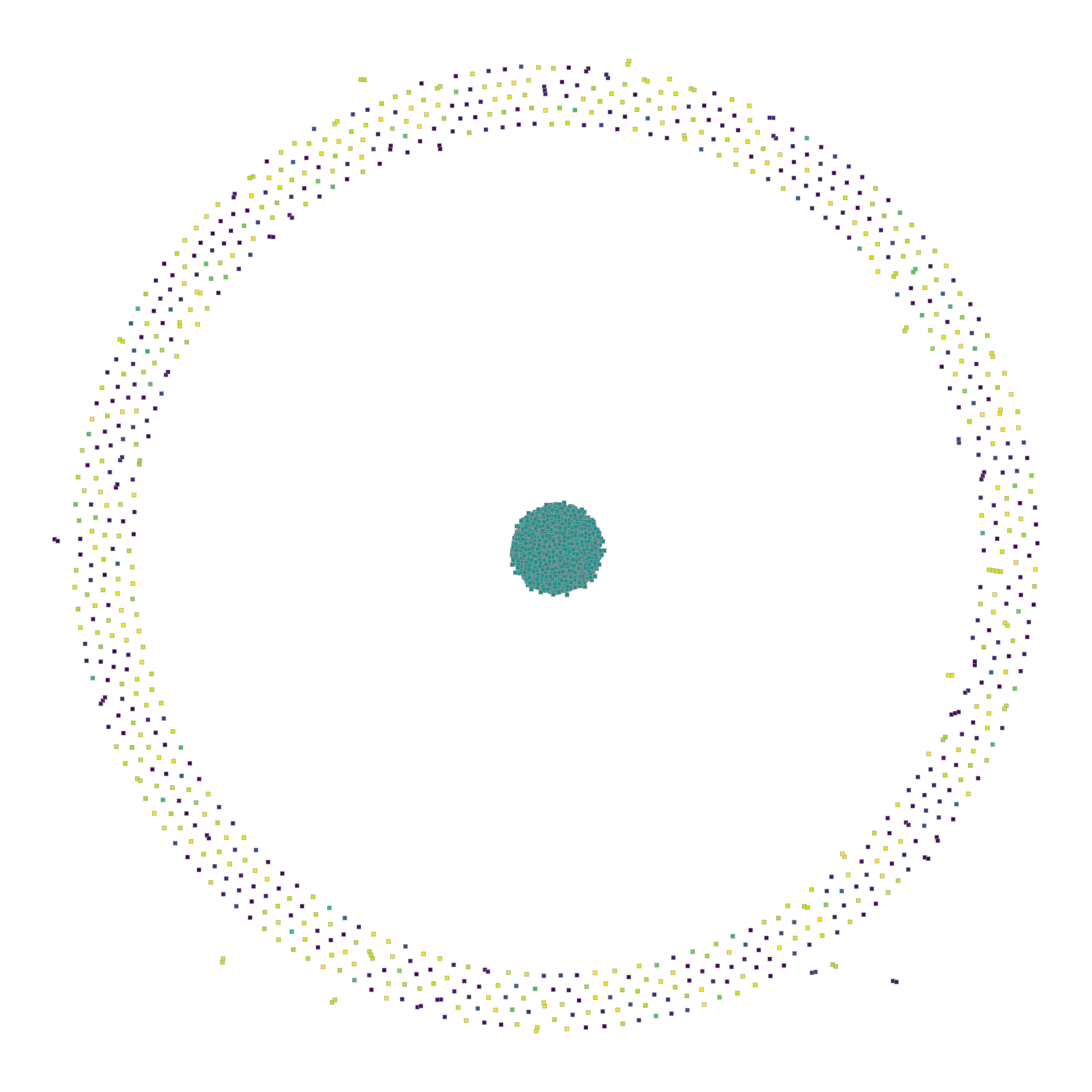}
            \caption{\label{fig:ba_graphs}
                Each panel shows the graph of the neighborhood relationship according to Eq.~\eqref{eq:neighbors},
                i.e., not all edges of the underlying BA graph but only those within confidence bounds of
                $\varepsilon = 0.14$ with $N=16384$ agents at $t=100$, $t=1000$, $t=9000$, $t=\infty$
                (from left to right). The color of the points representing the agents indicate their current opinion.
                The bridge of the mediating agents is very well visible pulling the two large clusters together.
                An animation of this process is available in the Supplemental Material \cite{supplemental}.
            }
        \end{figure*}

        Figure~\ref{fig:ba_graphs} shows four snapshots at different stages of the dynamics of a single realization of the BA network in the $\varepsilon$ region where the system changes from polarization to consensus with
        the clearly visible aforementioned bridges (the full animation, but on a static network layout, is available in the Supplemental Material \cite{supplemental}). For the sake of clarity, we also show in the right panel of Fig.~\ref{fig:ba_detailed} the complete time evolution of randomly selected agents of the same network, that shows how different clusters merge due to intermediate opinion agents (in color in the figure). As mentioned above, these agents in the bridges are so rare that they were not present in the random sample and had been added once they were identified in the network. However, bridges themselves occur in most
        realizations close to the transition region, which we checked using the betweenness centrality \cite{freeman1977set}, as shown in Sec.~VII of the Supplemental Material \cite{supplemental}.

        These bridges, which are the central mechanism enabling the consensus for very low $\varepsilon$ values can not exist in the complete network, since a group of agents only seeing one of the large clusters will be connected to all the agents of it and its opinion will very quickly join that of the cluster, thus preventing the interaction with the rest of the society.

        Finally, agents in a lattice have a small, regular neighborhood where they can potentially find other agents within their confidence interval. During the evolution, different clusters of connected agents holding similar opinions or \emph{domains} grow, as the magnetic domains in a phase transition of a magnetic system. If an agent on the \emph{domain wall} holds an opinion compatible with both domains, it can slowly lead them to merge. But as $\varepsilon$ decreases, this is less frequent and the system splits in clusters with opinions differing in at least $\varepsilon$. The square lattice facilitates a visualization of the evolution of this dynamics and one can see that neighborhoods mostly assume one of two opinions, which are outside
        of the confidence range of each other (and some isolated agents with extreme opinions). The animation showing this process can be found in the Supplemental Material \cite{supplemental}.

        This mechanism differs from the one described for random networks, as the lattice's topology does not allow the formation of relatively long bridges.
        As a consequence, when $\varepsilon$ decreases the fragmented state remains and $\varepsilon$ remains finite in the thermodynamic limit.

        We observe that the existence of consensus at very low confidences is
        inherently connected to the existence of a slow dynamics. This phenomenon seems characteristic of the HK model: it has also been observed in an heterogeneous version of the model \cite{schawe2020open}, which includes agents with different values $\varepsilon_i$ of confidence. In this case it was observed that when very closed minded agents (with low $\varepsilon_i$) were mixed with others with intermediate values of $\varepsilon_i$ consensus was reached due to a very slow dynamics. On the contrary the inclusion of more open minded agents accelerates the dynamics but destroys consensus.

    \section{Discussion}
    \label{sec:discussion}
        We have studied the dynamics the HK model in networked societies with different topologies ranging from lattices to random networks. For comparison, we also revisit the original model in the mixed population by performing detailed simulations by means of a fast algorithm~\cite{schawe2020open} which allows to reach very large sizes. In particular, we could obtain a very precise estimate of the confidence threshold.

        Choosing the normalized average size of the largest opinion cluster, as the order parameter that detects the transition to consensus, at a difference with the choice made in Ref.~\cite{fortunato2005consensus}, we are able to show that consensus is reached in networks at confidence values that are below the threshold for mixed populations. Moreover when comparing the HK dynamics in lattices (either pure or perturbed) with purely random networks, we observe that for the latter, the threshold for consensus shifts to zero in the thermodynamic limit, and converging times dramatically increase, meaning that consensus will be reached in a random network even at vanishing confidences, provided the time of the evolution is sufficiently long. Similarly, the fact that the introduction of disorder enhances consensus has been also observed in a study of the Axelrod model in small-world and scale free networks~\cite{klemm2003nonequilibrium}.
        With this order parameter we can also show that any network structure precludes the bifurcation phenomenon observed in the mixed population.

        Interestingly, the vanishing of the critical threshold is not a consequence of the degree distribution, as shown by the fact that it is present in both the ER and BA networks which are characterized by very different degree distributions (Poisson and scale free, respectively). A very detailed study of the temporal evolution of the agents' opinions, reveals that this effect depends on the existence of long bridges connecting clusters with opinions which differ in more than $2 \varepsilon$. After a large number of interactions these bridges succeed in bringing the two clusters to collapse.

        On the other hand, for two dimensional lattices, both regular and perturbed, our finite size scaling results show that the critical confidence threshold remains finite and we could obtain precise estimates for their values. We also show that the cluster size distribution at $\varepsilon_c$ follows a power law, a characteristic of a critical transition, with an Fisher exponent $\tau \sim 2$.

        However, the relation with percolation observed, for instance in the \textit{non consensus model} which can be mapped into invasive percolation~\cite{non_consensus}, is not straightforward for the HK model, due to the bounded confidence which is a constraint to the diffusion process. In fact, while the percolation threshold is zero for scale free networks~\cite{percolation_sf}, and finite for the ER networks~\cite{newman_book} both networks show a consensus threshold that shifts to zero as the size increases, due to the existence of bridges as we have seen.

        The striking qualitative difference in the tendency of $\varepsilon_c$ when studying the HK dynamics using $P_{u}$ as in Ref.~\cite{fortunato2005consensus} or $\avg{S}$ as here, comes from the fact that due to the relatively low mean number of neighbors found in networks, there may be agents with extreme initial opinions who do not interact at all, precluding unanimity, except for the trivial case of $\varepsilon \ge 0.5$, because at this point, any agent can interact with those holding the central consensus opinion of $x=0.5$.

        The other famous, usually better studied, bounded confidence model is the Deffuant model \cite{deffuant2000mixing}. In this case pairwise interactions are considered and the dyamics is \emph{sequential}, i.e., only two randomly drawn agents are updated at a time. Usually the two models present the same qualitative behavior: The Deffuant model shows a transition to consensus at a slightly higher critical confidence, its transition to unanimity is also at $\varepsilon = 0.5$ and $\avg{S}(\varepsilon)$ shows a similar bifurcation pattern. However, the sequential pairwise update prevents the formation of bridges, which are at the origin of the observed consensus in random networks for extremely low values of confidence in the HK model.
        We tested this prediction with simulations of the Deffuant dynamics, whose results are shown in Sec.~VIII of the Supplemental Material \cite{supplemental}.
        Indeed, our simulations of the Deffuant model on networks show that the transition to consensus happens at finite values of the confidence $\varepsilon$ also in random networks.
        On the other hand, same as the HK, the transition point is shifted to lower values of $\varepsilon$, therefore we conclude that the effect of sparse networks fostering consensus is robust.
        The raw data for the Deffuant model is available at \cite{rawDataDW}.

        At first sight our results might seem completely at odds with what it is observed in real social networks. Indeed, most social networks are very large and show a degree distribution that is reminiscent of scale free networks. However consensus is far from being always reached in social networks, where even extreme phenomena as communication bubbles and echo chambers are observed. In fact, many differences that can be observed between the HK stylized model and opinion dynamics processes in the real world such as the inclusion of idiosyncratic properties leading to agents' heterogeneity, or more complex diffusion dynamics rules, including other elements of Social Influence Theory, beyond conformity like minority influence, reactance~\cite{frager1970conformity}, obedience~\cite{blass1999milgram}, etc.
        Also some assumptions of our model might be violated in realistic systems: We studied the final states, which are only reached after very long convergence times, such that they might never be reached in real societies, where circumstances change on a faster timescale, which would suggest that we can only observe transient states of polarized and fragmented societies. Furthermore, we assumed that the network where opinion diffuses is \textit{connected}. In social networks the connections between agents are mediated by algorithms that select which content will be addressed to the agent according to its profile leading to an effective disconnection of the network which hinders an efficient diffusion of the opinion. Similarly, the studied models do not include correlation between the network and the opinions, while in reality special interest groups or forums establish new connection between like minded people.

    \section{Conclusions}
    \label{sec:conclusions}
        This work shows that the bounded confidence Hegselmann-Krause opinion dynamics model yields non trivial results in network topologies.

        In particular, we show that randomness enhances consensus, which is at odds with a previous result~\cite{fortunato2005consensus} but in agreement with the results observed for the Axelrod model~\cite{klemm2003nonequilibrium} in networks.

        We show that the vanishing of the consensus threshold in the thermodynamic limit, observed in purely random networks is due to the existence of bridges, containing several nodes, between clusters of very different opinions that could not interact otherwise. The nodes in the bridge, through successive interactions allow the clusters to merge, hence the dramatic increase of the convergence time. This mechanism is absent from lattices either pure of perturbed by a WS algorithm with low rewiring probability.

        Although the HK model is a stylized approximation to the dynamics of opinion formation and diffusion in the real world, the considered interactions are present---among others---in real systems. One aspect of social interactions in the real world is that they are layered, each of us holds social interactions in different environments (household, work, leisure activities, on-line media, etc.). Given the importance of the type of connectivity shown here, our next step would be to investigate the behavior of this model in a multiplex network.

    \section*{Acknowledgments}
        The authors thank R\'{e}mi Perrier for suggesting improvements to the manuscript.
        The authors also acknowledge the OpLaDyn grant obtained in the 4th round
        of the Trans-Atlantic Platform Digging into Data Challenge (2016-147 ANR OPLADYN TAP-DD2016)
        and Labex MME-DII (Grant No. ANR reference 11-LABEX-0023).

    \bibliography{lit}

\end{document}

\title{Supplementary Material for: \\
    The bridges to consensus: Network effects in a bounded confidence opinion dynamics model}

\author{Hendrik Schawe}
    \email{hendrik.schawe@cyu.fr}
    \affiliation{Laboratoire de Physique Th\'{e}orique et Mod\'{e}lisation, UMR-8089 CNRS, CY Cergy Paris Universit\'{e}, France}
    \author{Sylvain Fontaine}
    \affiliation{Laboratoire de Physique Th\'{e}orique et Mod\'{e}lisation, UMR-8089 CNRS, CY Cergy Paris Universit\'{e}, France}
    \author{Laura Hern\'{a}ndez}
    \email{laura.hernandez@cyu.fr}
    \affiliation{Laboratoire de Physique Th\'{e}orique et Mod\'{e}lisation, UMR-8089 CNRS, CY Cergy Paris Universit\'{e}, France}

    \date{\today}

    \maketitle

    \tableofcontents
    
    \section{Supplementary Videos}
    \label{videos}
        We provide 3 animations that illustrate the descriptive text of the
        main manuscript and also highlight the qualitative difference of the dynamics on lattices
        and random graphs. In all of them the opinion of the agent is coded by a color scale going from yellow to dark blue, with the intermediate equilibrium opinion coded by a dark cyan color.
        \begin{itemize}
            \item \texttt{lat3\_eps015.mp4} shows the
                evolution of the agents' opinions (same system as shown in Fig.~3(c) of the main manuscript)
                on a square lattice with third neighbors at $\varepsilon = 0.15$.
            \item \texttt{lat4\_eps015.mp4} shows the
                evolution of the agents' opinions (same system as shown in Fig.~3(d) of the main manuscript)
                on a square lattice with fourth neighbors at $\varepsilon = 0.15$.
            \item \texttt{ba10\_eps015.mp4} shows the
                evolution of the agents' opinions (same system as shown in Fig.~3(d) of the main manuscript)
                on a BA network with mean degree of $\avg{k} = 10$ at $\varepsilon = 0.15$.
        \end{itemize}
        The videos should facilitate easier understanding of the descriptive text of the
        main manuscript and also highlight the qualitative difference of the dynamics on lattices
        and random graphs.

    \section{Fully connected, Cluster size distribution}
    \label{fully_connected}
\begin{figure}[htb]
            \centering
            \includegraphics[scale=1]{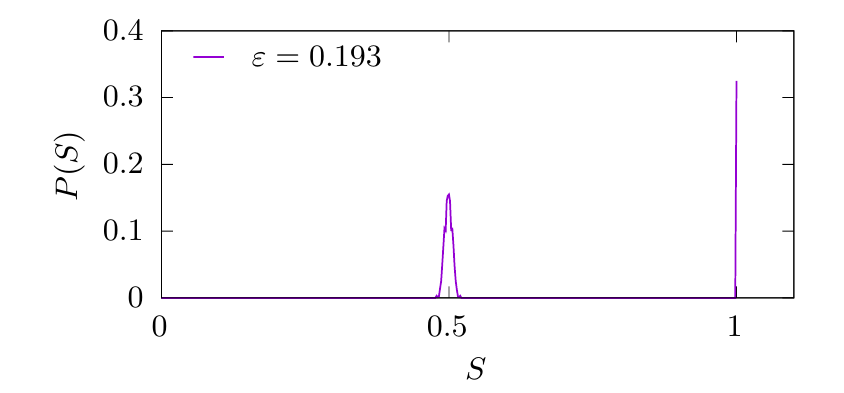}
            \caption{\label{fig:fully_dist}
                Cluster size distribution in the fully connected society at $\varepsilon = 0.193$
                close to the critical threshold $\varepsilon = 0.1926(5)$ for $N = 262144$.
            }
        \end{figure}

        Figure 5 of the main text shows for the case of the square lattice with third nearest
        neighbors that the cluster size distribution $P(S)$ follows a power-law. Here, in Fig.~\ref{fig:fully_dist}
        we show the cluster size distribution at the threshold for the case of a mixed population,
        which shows a coexistence
        of two values corresponding to unanimous and polarized states. The broad distribution
        found for  lattices is therefore a fundamental difference to the behavior of the
        well mixed case.

    \section{Square Lattice: first, second and fourth neighbors}
    \label{square}
        The main manuscript shows the behavior for square lattices including connections to third
        neighbors. This selection was chosen, since the connectivity of $\avg{k} = 12$ for this
        case is close to the connectivity of $\avg{k} = 10$ of the random networks also studied.

\begin{figure}[htb]
            \centering
            \includegraphics[scale=1]{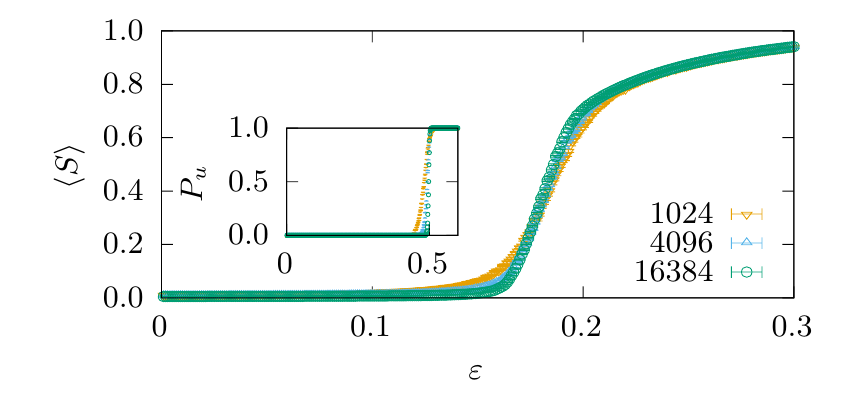}
            \includegraphics[scale=1]{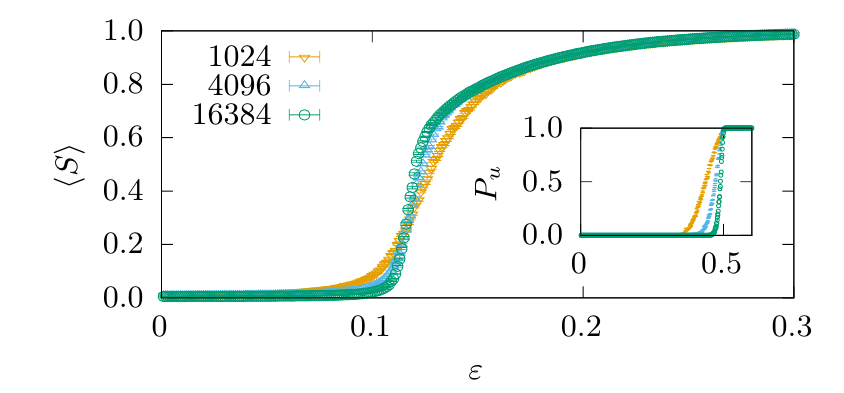}
            \caption{\label{fig:lattice:1}
                Mean size of the largest cluster $\avg{S}$ as a function of the confidence
                $\varepsilon$ for multiple sizes for the square lattice with left: first nearest neighbors,
                right: second nearest neighbors.
                The insets shows the probability of unanimity on a different $\varepsilon$ axis.
            }
        \end{figure}

\begin{figure}[htb]
            \centering
            \includegraphics[scale=1]{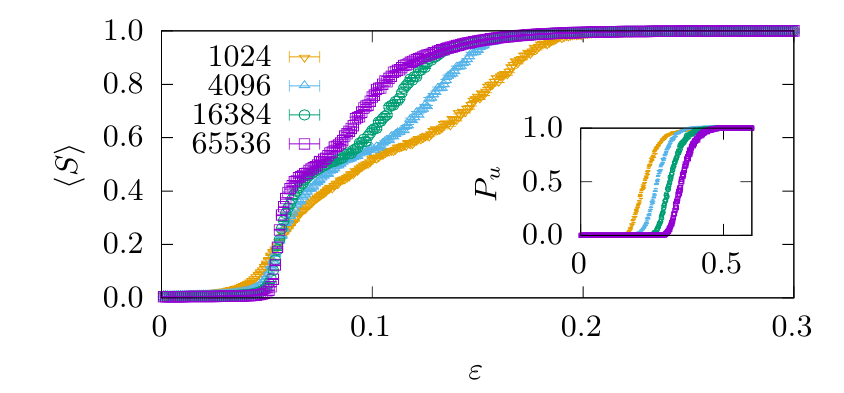}
            \includegraphics[scale=1]{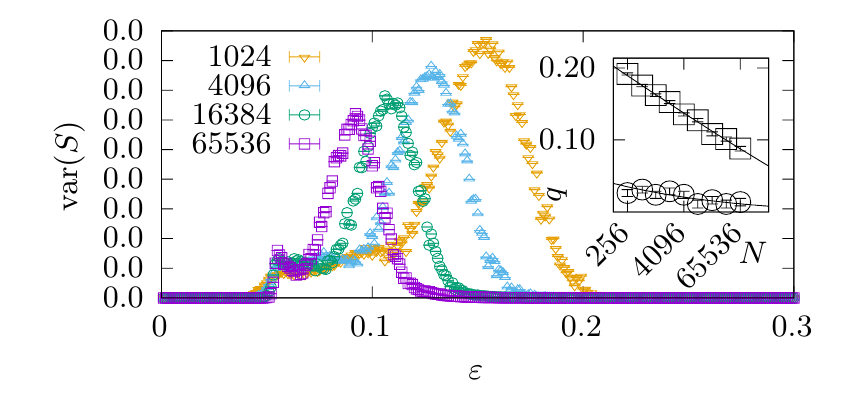}
            \caption{\label{fig:lattice:4}
                Left panel: Mean size of the largest cluster $\avg{S}$ as a function of the confidence
                $\varepsilon$ for multiple sizes of the square lattice with fourth nearest neighbors.
                The inset shows the probability of unanimity on a different $\varepsilon$ axis.
                Right panel: Variance of the size of the largest cluster $\mathrm{var}(S)$ for the square lattice
                with up to fourth nearest neighbors. The right peaks converge fast to the
                left (squares in the inset), while the left peaks converge towards the
                transition point (circles in the inset). The inset shows the positions
                of both peaks in a log-log plot and extrapolations to a power law with offset.
            }
        \end{figure}

        However, qualitatively the same behavior occurs when looking at standard $\avg{k} = 4$
        square lattices, second neighbor square lattices with $\avg{k} = 8$ (both in Fig.~\ref{fig:lattice:1})
        and fourth neighbor lattices $\avg{k} = 20$ (in Fig.~\ref{fig:lattice:4}).
        In all cases the transition point stays finite, estimates
        for its value are listed in table I of the main manuscript, except for the second nearest
        neighbor case, where we estimate $\varepsilon_\times= 0.1155(3)$.

        Note that the first and second neighbor cases do not show the plateau, which was visible
        for third nearest neighbors.
        In the fourth nearest neighbor case, the plateau becomes a bit
        more pronounced and therefore a bit more similar to the pattern
        from the fully connected case. A pattern we also observed for random networks
        of different connectivity in Fig.~9 of the main manuscript.
        Observing the behavior for increasing sizes $N$, we
        are very confident, that it will vanish again in the limit of large systems.
        This is supported again by the extrapolation of the corresponding peak positions, where the
        right flank seems to move like a power-law to $\varepsilon \approx 0$, while the left flank
        shows, analogue to the $\avg{k} = 12$ case, a convergence towards $\varepsilon_\wedge = 0.049(26)$,
        which is again compatible with the transition point obtained by the
        the crossing points $\varepsilon_\times = 0.0548(7)$.
        On the other hand, the unanimity
        parameter, shown in the inset of the top panel, has due to the rather large connectivity
        difficulties to determine the
        transition point despite the extremely large systems we simulated---also a
        well known effect \cite{fortunato2005consensus}.

    \section{Perturbed Square Lattice, third neighbors}
    \label{perturbed}
\begin{figure}[htb]
            \centering
            \includegraphics[scale=1]{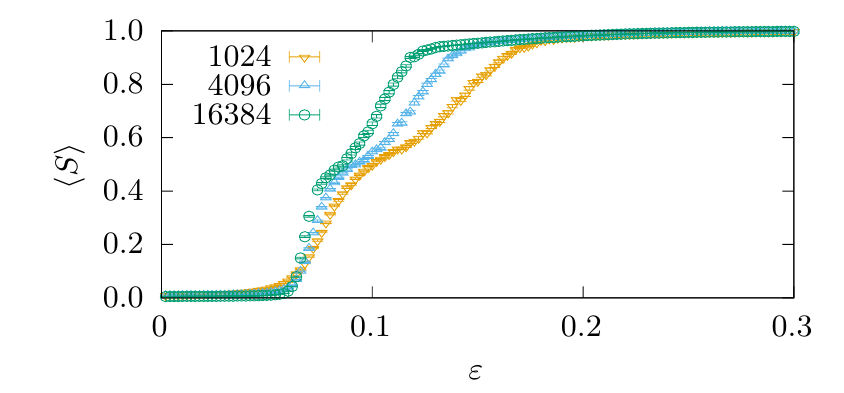}
            \includegraphics[scale=1]{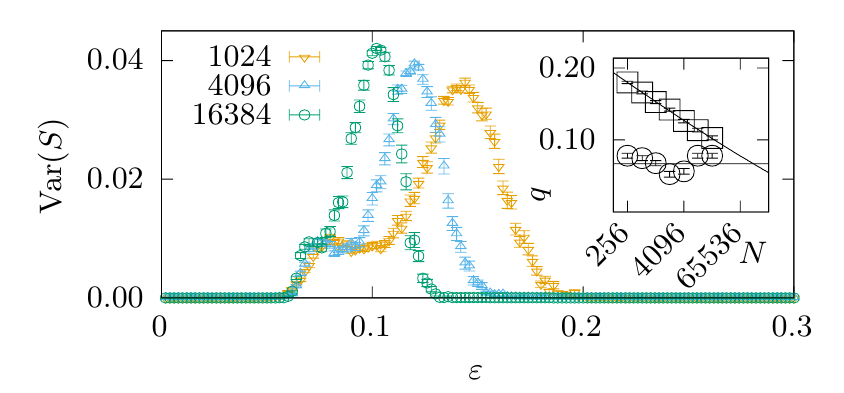}
            \caption{\label{fig:ws:12}
                Left panel: Mean size of the largest cluster $\avg{S}$ as a function of the confidence
                $\varepsilon$ for multiple sizes for the perturbed lattice, $p=0.01$, with third neighbors.
                Right panel: Variance of the size of the largest cluster. The inset shows the extrapolation
                of the positions of the two peaks.
            }
        \end{figure}

        We present results for the perturbed lattice, where $p=0.01$ of all edges are randomly
        rewired in Section III.B.2.~of the main manuscript. Figure \ref{fig:ws:12} shows
        the data used to reach the conclusions presented in the main manuscript. It shows
        the same characteristics as the pure lattice with third neighbors: the polarization
        plateau, the two corresponding peaks in the variance in the second panel and the
        same pattern, where the left peak in the variance converges towards a finite value
        and the right peak decaying like a pure power law, until they merge.

    \section{Barab\'asi Albert, $\avg{k} = 4$}
        \label{sf}
        Similar to the lattice case (cf.~Fig.~\ref{fig:lattice:1}), the plateau of polarization does not occur
        for low values of the mean degree $\avg{k}$ also in the BA ensemble. But still
        all characteristic differences of the two ensembles in relation to the HK model remain:
        The lattice case shows crossing, while the BA does not. And an extrapolation of the positions
        of the maxima in the variance clearly shows a power-law decay towards zero. To quantify this,
        we fitted a power-law with an offset according to Eq.~(4) (listed in table I of the main manuscript),
        which is compatible with an offset of $0$.

        \begin{figure}[htb]
            \centering
            \includegraphics[scale=1]{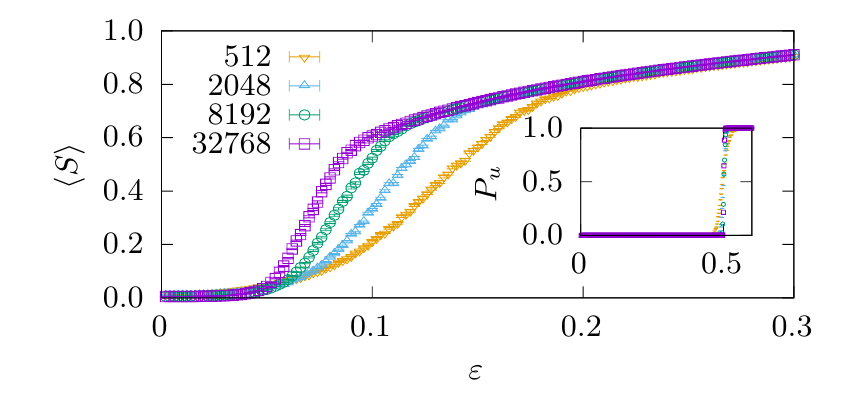}
            \includegraphics[scale=1]{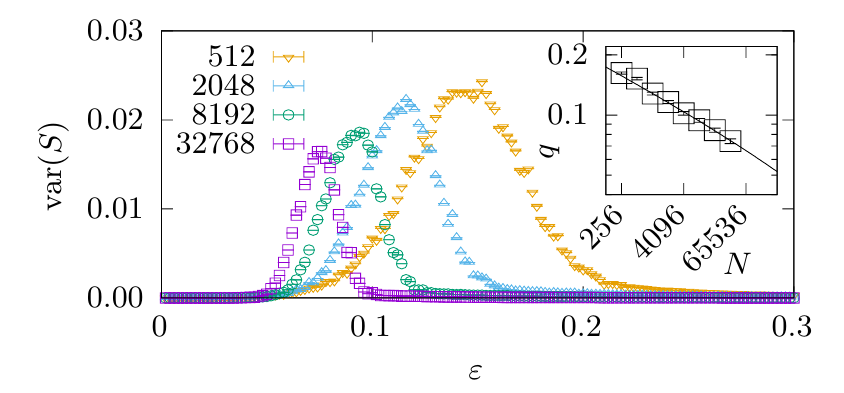}
            \caption{\label{fig:ba:4}
                Left panel: Mean size of the largest cluster $\avg{S}$ as a function of the confidence
                $\varepsilon$ for multiple sizes of the BA with mean degree of $\avg{k} = 4$.
                The inset shows the probability of unanimity on a different $\varepsilon$ axis.
                Right panel: Variance of the size of the largest cluster $\mathrm{var}(S)$ for
                the BA with mean degree of $\avg{k} = 4$. The peaks converge fast to the
                left (squares in the inset). The inset shows the positions
                of the peaks in a log-log plot and extrapolations to a power law with offset,
                where the fitted value of the offset is within statistical error compatible with zero.
            }
        \end{figure}

    \section{Erd\H{o}s R\'enyi, $\avg{k} = 10$}
    \label{ER}
        Figure 6 and 8 of the main manuscript show the entropy $\sigma$ as a function of
        the confidence $\varepsilon$, to show that the lattice case exhibits a narrowing peak,
        while the BA case exhibits a shifting plateau. Here we show the same for the ER case,
        which shows, in accordance with the results stated in the main manuscript a shifting
        plateau similar to the BA case.

        \begin{figure}[htb]
            \centering
            \includegraphics[scale=1]{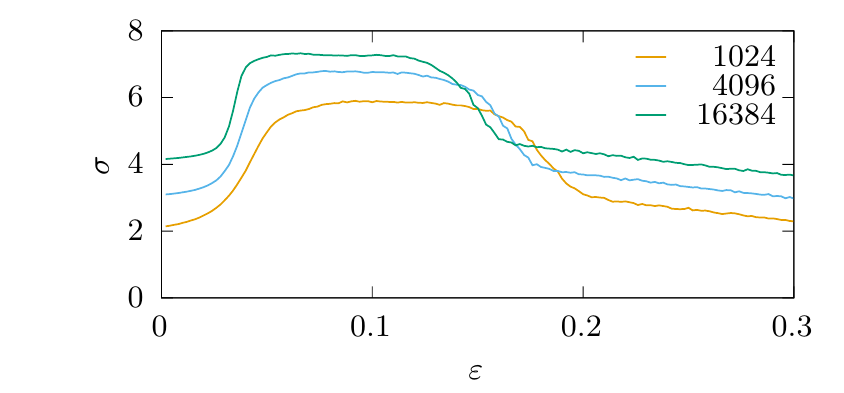}
            \caption{\label{fig:er:10}
                Entropy $\sigma$ of the cluster-size distribution for the
                Erd\H{o}s-R\'enyi ensemble with a mean degree of $\avg{k} = 10$.
                It does not show a sharp peak at finite values of $\varepsilon$. Its maximum
                is stretched over a plateau, which moves continuously towards
                0, with is consistent with the before estimated transition point
                at 0.
            }
        \end{figure}

    \section{Measuring the Bridges}
    \label{bridge}
        We showed the existence of the bridges directly at a representative example in the main manuscript. As a quantitative proxy for the presence of bridges at different values of the confidence $\varepsilon$, we propose to look at the maximum betweenness centrality $g$ \cite{freeman1977set} reached by any node over the full dynamics. The betweenness centrality is defined per node as the number of shortest paths between all pairs of nodes which pass through it, divided by the number of pairs.
        In the presence of bridges between two roughly equal clusters, we would expect to observe a maximum betweenness centrality of $g \approx 0.5$, since the shortest path between every second pair would have to traverse the bridge.
        Note that while the underlying topology is static, the effective interaction network is dynamic since it depends on the time dependent opinions of neighboring agents.
        For computational reasons, we do not calculate the betweenness centrality after every sweep, but in exponentially increasing intervals, such that our measurements will underestimate the actual maximum. Also we do not calculate the exact betweenness centrality but an approximation obtained from 1000 random sampled pairs instead of all pairs. This approximated value is then averaged over 100 different realizations for each value of $\varepsilon$.

        \begin{figure}[htb]
            \centering
            \includegraphics[scale=1]{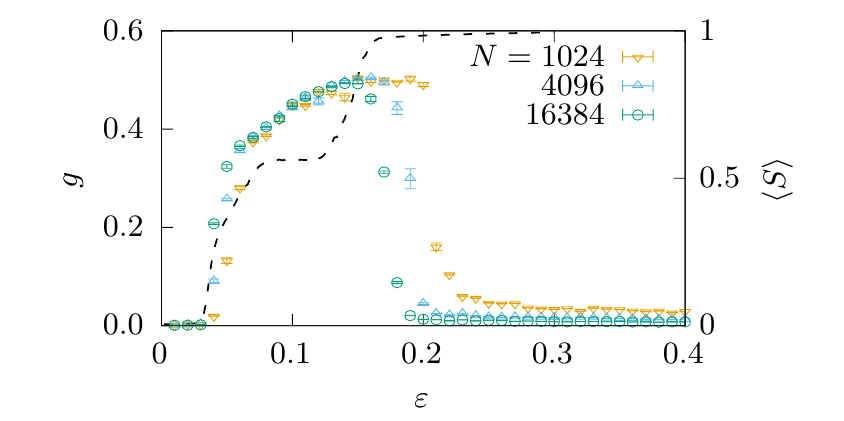}
            \includegraphics[scale=1]{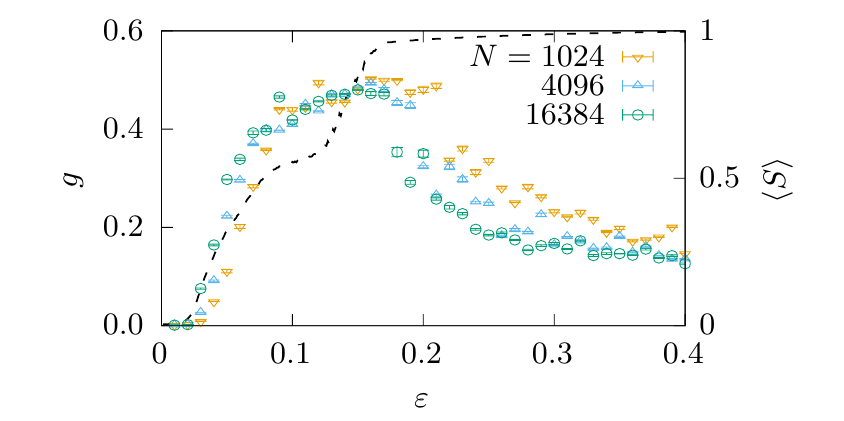}
            \caption{\label{fig:betweenness}
                The betweenness centrality $g$ as a function of $\varepsilon$ for the ER (left) and BA (right) at a connectivity of $c=10$. The dashed line shows the order parameter $\avg{S}$ for $N=16384$ as a reference.
            }
        \end{figure}

        Figure \ref{fig:betweenness} shows the betweenness centralities for ER and BA both with connectivity $c=10$. Both show a clear maximum in the critical region ($\approx$ the region over which the slope of the dashed line goes from polarization to consensus) with values very close to $g = 0.5$, which is a very clear sign that in this parameter region a bridge existed in almost all realizations at one point in the simulation. Also, for larger system sizes bridges appear at lower values of $\varepsilon$, enabling the formation of consensus at lower confidences and supporting the main result of the main manuscript: for large enough systems consensus is obtained for arbitrarily low confidence in random networks. At larger values of $\varepsilon$ bridges disappear, because the dynamics changes to a single cluster dominating from the beginning. Also, as expected, due to the presence of hubs, which connect many nodes, the BA shows generally a higher betweenness centrality than the ER.

    \section{The Deffuant Model}
    \label{deffuant}

        For completeness, we simulated the other influential bounded confidence model: The Deffuant model. Its definition is similar to the HK case, but the update happens sequentially between random pairs of agents
        $(i, j)$ according to the rules:

        \begin{align}
            x_i(t+1) = \begin{cases}
                \mu x_i(t) + (1-\mu) x_j(t), & \text{if } |x_i - x_j| < \varepsilon\\
                x_i(t), & \text{otherwise}
            \end{cases}\\
            x_j(t+1) = \begin{cases}
                \mu x_j(t) + (1-\mu) x_i(t), & \text{if } |x_i - x_j| < \varepsilon\\
                x_j(t), & \text{otherwise}.
            \end{cases}
        \end{align}

        Here, we set $\mu = 0.5$, i.e., both agents reach their average opinion after one interaction. This choice
        should lead to fast convergence times while not changing the final configuration strongly.

        Besides the changed update rule, we keep all other parameters fixed: the same uniformly distributed initial opinions, the same convergence criterion and a subset of topologies.

        We observe that on sparse topologies the Deffuant model generally takes a longer time to converge, such that we can not study the same sizes as we did for the HK case in the main part of the manuscript. Especially, even for the shown sizes, not all simulations did finish, thus we excluded all datapoints for the corresponding parameter set to avoid selection bias. The raw data for the available parameter sets are available at \cite{rawDataDW}.

\begin{figure}[htb]
            \centering
            \includegraphics[scale=1]{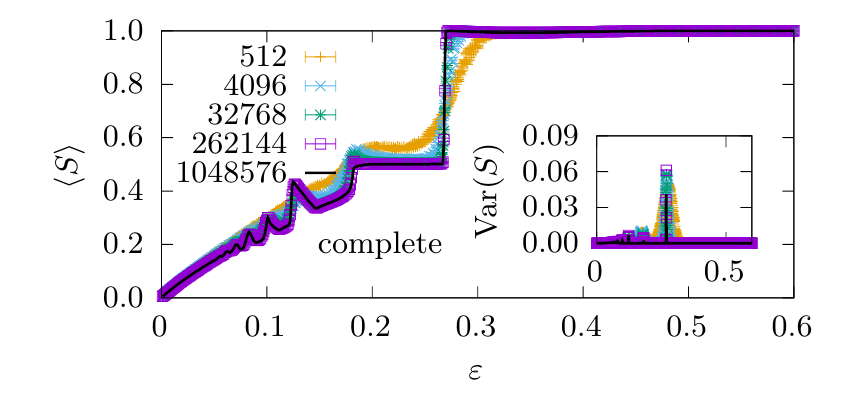}
            \includegraphics[scale=1]{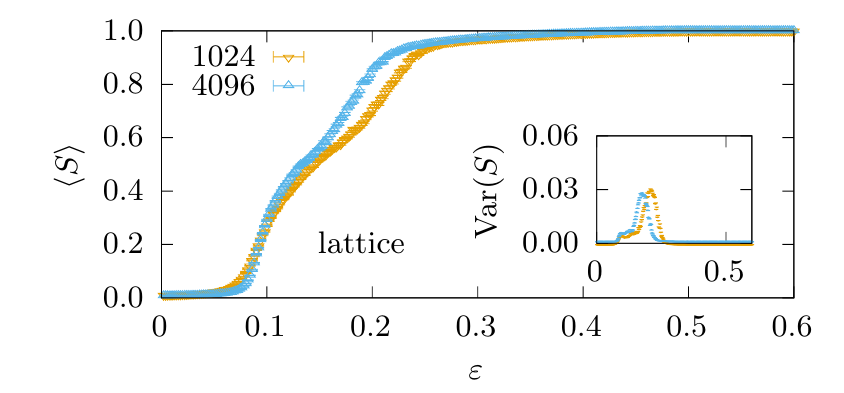}
            \includegraphics[scale=1]{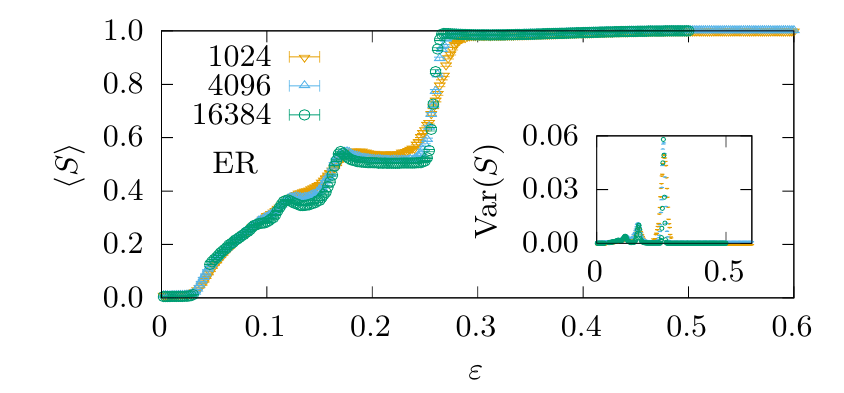}
            \includegraphics[scale=1]{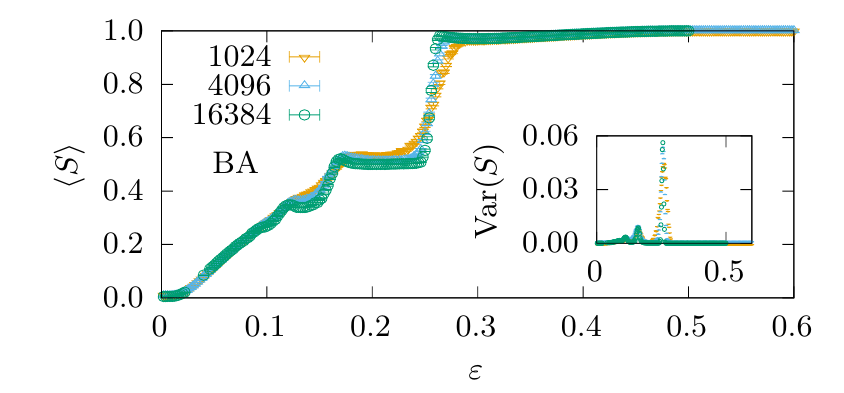}
            \caption{\label{fig:deffuant}
                Mean size of the largest cluster $\avg{S}$ as a function of the confidence $\varepsilon$ for the Deffuant model on different topologies: the complete graph, a square lattice with third nearest neighbors, an ER graph with connectivity of $c=10$ and a BA graph with connectivity $c=10$. Each datapoint is an average over $1000$ random realizations.
            }
        \end{figure}

        In Figure~\ref{fig:deffuant} we show the order parameter $\avg{S(\varepsilon)}$ for different topologies. The well studied case of the complete graph is pictured for reference. Indeed it shows the well known bifurcation phenomenon and a transition to consensus at $\varepsilon_c = 0.2691(2)$, which we measured from the crossing of multiple system sizes as described in the main manuscript.

        Also the square lattice with up to third nearest neighbors behaves similarly to the HK case: The variance in the inset shows two peaks, one going to the left and one staying at the critical point. The critical point estimated by crossing, is at $\varepsilon_c = 0.096(1)$ far lower than in the fully connected case, which means that the general effect that sparse networks foster consensus seems to be robust against this modification of the model.

        Most interesting are the two random graph ensembles. Unlike for the HK dynamics, for the Deffuant model the shape of $\avg{S(\varepsilon)}$ for random graphs is almost unchanged with respect to the corresponding fully mixed population case. The transition to consensus still happens at a clearly defined finite value, which is only very slightly lower than in the complete graph at $\varepsilon_c = 0.257(6)$ for ER and $\varepsilon_c = 0.253(7)$ for BA. Especially the bridges can not exist over many sweeps in a system with pairwise sequential updates: a single update will move a member of the bridge to the same position as another member and directly destroy the bridge.

    \bibliography{lit}